\documentclass[showkeys,eqsecnum,showpacs,prd]{revtex4}
\usepackage[dvips]{graphicx}
\usepackage{color}
\usepackage{amsmath}
\begin{document}

\title{Partition Function of Spacetime}

\author{Jarmo M\"akel\"a} 

\email[Electronic address: ]{jarmo.makela@puv.fi}  
\affiliation{Vaasa University of Applied Sciences, Wolffintie 30, 65200 Vaasa, Finland}

\begin{abstract}
We consider a microscopic model of spacetime, where spacetime is assumed to be a specific graph with Planck size
quantum black holes on its vertices. As a thermodynamical system under consideration we take a certain uniformly
accelerating, spacelike two-surface of spacetime which we call, for the sake of brevity and simplicity, as 
{\it acceleration surface.} Using our model we manage to obtain an explicit and surprisingly simple expression for 
the partition
function of an acceleration surface. Our partition function implies, among other things, the Unruh and the Hawking
effects. It turns out that the Unruh and the Hawking effects are consequences of a specific phase transition, which
takes place in spacetime, when the temperature of spacetime equals, from the point of view of an observer at rest
with respect to an acceleration surface, to the Unruh temperature measured by that observer. When constructing the
partition function of an acceleration surface we are forced to introduce a quantity which plays the role of 
thermal energy of the surface. An interpretation of that quantity as energy in a normal manner yields Einstein's 
field equation with a vanishing cosmological constant for general matter fields.

\end{abstract}


\maketitle

\section{Introduction}
 
  Is spacetime made of some elementary constituents in the same way as matter is made of atoms? This is one of the 
central questions of the so called {\it emergent gravity}, which views gravity as an emergent, instead of a 
fundamental property of spacetime. According to some ideas of emergent gravity, which have gained increasing 
popularity during some recent years, gravitation appears at macroscopic length scales as a consequence of the
properties of a some still unknown substructure of spacetime in the same way as, say, an elasticity of a solid 
body is a consequence of the properties of its atoms. \cite{yksi} If the aims of emergent gravity were realized, one
should be able to obtain all of the "hard facts" of gravity as such as we know them today, together with some
possible new predictions, from an appropriate microscopic model of spacetime. Those hard facts include, among
other things, Einstein's classical general relativity with all of its consequences, together with the Unruh and 
the Hawking effects. An ability to predict general relativity, as well as the Unruh and the Hawking effects, at
macroscopic length scales acts as a crucial test for any viable microscopic model of spacetime.

   When one goes over from a microscopic to a macroscopic description of any system, the key role is played
by the {\it partition function}
\begin{equation}
Z(\beta) := \sum_n g(E_n)e^{-\beta E_n}
\end{equation}
of the system. If we know the partition function of a system, we may calculate the macroscopic quantities
relevant to the system as functions of its inverse temperature $\beta$ in a very simple manner. For instance,
the average total energy of the system is
\begin{equation}
E(\beta) = -\frac{\partial}{\partial\beta}\ln Z(\beta),
\end{equation}
and its entropy is, in natural units:
\begin{equation}
S(\beta) = \beta E(\beta) + \ln Z(\beta).
\end{equation}
In Eq.(1.1) $E_n$ denotes the possible energy eigenvalues of a system, and $g(E_n)$ is the number of degenerate states 
corresponding to the same energy eigenvalue $E_n$. When one attempts to obtain the macroscopic propeties of gravity
from an appropriate microscopic model of spacetime, one must first calculate the partition function of spacetime.
The properties of the partition function should then imply the properties of gravity at macroscopic length scales.

   In this paper we consider a specific microscopic model of spacetime. Using our model we manage to obtain an 
explicit
- and surprisingly simple - expression for the partition function of spacetime. Our partition function implies,
among other things, the Hawking and the Unruh effects, together with a formula very similar to the 
Bekenstein-Hawking entropy law for black holes. According to our model the Hawking and the Unruh effects are
consequences of a certain type of {\it phase transition}, which takes place in spacetime. During this 
phase transition the fundamental constituents of spacetime jump from a one quantum state to another in a very
specific way. When calculating the partition function we are forced to introduce a concept, which plays the role
of thermal energy in our model. An interpretation of that concept as energy in a normal manner implies Einstein's
field equation with a vanishing cosmological constant. 

     This paper is a continuation to a series of papers, where {\it Planck size quantum black holes} were used as
the fundamental constituents of spacetime. \cite{kaksi, kolme, nelja} As in those papers, we model spacetime by a specific graph, where
black holes lie on the vertices. The only physical degree of freedom associated with an individual black hole is
its event horizon area, which is assumed to have a discrete spectrum with an equal spacing. More precisely, the
eigenvalues of the event horizon area are assumed to be of the form
\begin{equation}
A_n = (n + \frac{1}{2})32\pi\ell_{Pl}^2,
\end{equation}
where $n = 0, 1, 2,...$, and 
\begin{equation}
\ell_{Pl} := \sqrt{\frac{\hbar G}{c^3}}\approx 1.6 \times 10^{-35}m
\end{equation}
is the Planck length. A horizon area spectrum with an equal spacing for black holes was proposed by Bekenstein 
already in 1974. \cite{viisi} Since then Bekenstein's proposal has been recovered by several authors on various 
grounds, and it has been an object of wide and detailed investigations. \cite{kuusi} One of the key ideas of 
our model of spacetime is to reduce all properties of spacetime to the horizon area eigenstates of the 
Planck size quantum black holes constituting spacetime. \cite{kaksi}

      There are some indications that at the Planck scale of distances microscopic black holes might really play
some role in the structure of spacetime. For instance, if we close a particle inside a box with an edge length 
equal to the Planck length $\ell_{Pl}$, then Heisenberg's uncertainty principle implies that the momentum of the
particle has an uncertainty $\Delta p \sim \hbar/\ell_{Pl}$. In the ultrarelativistic limit the uncertainty 
$\Delta p$ in the momentum corresponds to the uncertainty $\Delta E\sim c\Delta p$ in the energy of the particle.
In other words, inside a box with edge length equal to $\ell_{Pl}$ we have closed a particle, which has the 
Planck energy 
\begin{equation}
E_{Pl} := \sqrt{\frac{\hbar c^5}{G}} \approx 2.0\times 10^9 J
\end{equation}
as an uncertainty in its energy. This energy, however, is enough to shrink the box into a black hole with a 
Schwarzschild radius equal to about one Planck length. So it seems that when probing spacetime at the Planck 
scale of distances one is likely to meet with Planck size black holes.

      The idea that spacetime might consist of tiny black holes is far from new. Somewhat related ideas have
been expressed, for instance, by Misner, Thorne and Wheeler in their book. \cite{seitseman} Unfortunately, the idea of spacetime 
being made of Planck size black holes has never been taken very far. This paper is a part of an ongoing project, 
where this idea is being explored systematically.

      The very first question one should always ask at the beginning of an investigation of a physical problem is: 
"What is the {\it system} under consideration?" For a solid state physicist, for instance, the system may be a 
piece of metal, and he may be interested to explain its macroscopic properties, such as specific heat and electric
conductivity, by means of the properties of its atoms, whereas an elementary particle physicist, in turn, may be 
interested in a system which consists of two elementary particles, and a particle which intermediates their 
mutual interactions.

       In this paper we propose that the system one should consider in gravitational physics at the macroscopic
length scales is the so called {\it acceleration surface} which was introduced in Ref. \cite{kolme}. To put it simply, an
acceleration surface may be described as a smooth, orientable, simply connected, spacelike two-surface of spacetime,
whose every point accelerates uniformly with a constant proper acceleration to the direction of a spacelike unit
normal vector field of the surface. Examples of acceleration surfaces include, among other things, a uniformly 
accelerating spacelike plane in flat Minkowski spacetime,
and the $t = constant$ slices of a timelike hypersurface, where $r = constant$ $(>2M)$ in Schwarzschild spacetime
equipped with the Schwarzschild coordinates. In our model acceleration surfaces are assumed to be made of Planck
size quantum black holes in a somewhat similar way as solids are made of atoms, and from the postulated properties
of those black holes we infer the macroscopic properties of acceleration surfaces.

       A detailed investigation of the geometrical and the dynamical properties of acceleration surfaces was
performed in Ref. \cite{kolme}. To make our presentation self-contained, we begin our considerations, in Section 2, by an
extensive review of those properties. One of the motivations for introducing the concept of acceleration surface
is Jacobson's remarkable discovery of the year 1995 that Einstein's field equation may be obtained from the 
thermodynamical relation $\delta Q = T\,dS$ and certain assumptions concerning the properties of the local past Rindler
horizon of an accelerating observer. \cite{kahdeksan} More precisely, Jacobson assumed that when matter flows through of a finite
part of the horizon, that part shrinks such that the entropy carried by the matter through the horizon is, in 
natural units, exactly one-quarter of the decrease in the area of that part. In other words, Jacobson managed
to show that Einstein's field equation may be obtained from certain thermodynamical assumptions concerning the
properties of Rindler horizons. Acceleration surfaces may be viewed as generalizations of the horizons of spacetime
in the sense that an arbitrary horizon of spacetime is always a limit of a certain acceleration surface, if we take 
the proper acceleration $a$ of that surface to infinity. One of the key questions is, whether Jacobson's 
thermodynamical derivation of Einstein's field equation may be extended for surfaces more general than Rindler
horizons, and acceleration surfaces provide an appropriate generalization.

     In Section 2 we present a mathematically precise definition of the concept of acceleration surface. When an
acceleration surface propagates in spacetime, its area may change, and we write the equations, exactly derived in 
Ref. \cite{kolme}, which tell how the changes in the area of an acceleration surface depend on the geometry of the 
underlying spacetime. A question of fundamental importance is, whether it is possible to associate a quantity
analogous to {\it thermal energy} with acceleration surfaces in any physically meaningful way. Motivated by the analogies
drawn from Newtonian gravity, Unruh effect, and the mass formula of black holes we argue that this is indeed 
possible. Additional motivation for defining the concept of thermal energy, or heat, for an acceleration surface
in a certain manner is provided by the fact that Einstein's field equation with a vanishing cosmological constant 
may be obtained from a very simple equation which describes the exchange of heat between an acceleration surface
and massless, non-interacting radiation flowing through that surface. Since that equation really implies, as we 
show in Section 2, Einstein's field equation with a vanishing cosmological constant, not only for radiation, 
but for general matter fields, we call that equation, in our model, as the "fundamental equation" of the 
thermodynamics of spacetime. The derivation of Einstein's field equation from our fundamental equation
provides a generalization of Jacobson's thermodynamical derivation.  

     In Section 3 we proceed from the macroscopic and classical description of acceleration surfaces 
performed in Section
2 to their microscopic and quantum-mechanical description. We pose four {\it independence-} and two 
{\it statistical} postulates for the black holes constituting the acceleration surfaces of spacetime. The 
independence postulates imply, among other things, that the possible eigenvalues of the area $A$ of an 
acceleration surface are of the form:
\begin{equation}
A = \alpha\ell_{Pl}^2(n_1 + n_2 + ... + n_N),
\end{equation}
where $\alpha$ is a pure number of the order of unity, and the non-negative integers $n_1, n_2,..., n_N$ are the
quantum numbers labelling the horizon area eigenstates of the $N$ individual black holes which constitute the 
surface. The statistical postulates, in turn, identify the microscopic and the macroscopic states of an acceleration
surface. The microscopic states of an acceleration surface are determined by the different combinations of the
non-vacuum ($n\ne 0$) horizon area eigenstates of the quantum black holes on the surface, and each microscopic
state yielding the same area eigenvalue of the surface corresponds to the same macroscpic state. When writing the 
partition function of an acceleration surface we identify the energy of the surface as the thermal energy defined
in Section 2. Our definition implies that there is a one-to-one relationship between the area- and the energy 
eigenvalues of a given acceleration surface, and therefore the number $g(E_n)$ of the degenerate states 
corresponding to the same energy eigenvalue $E_n$ of the surface equals to the number of microstates corresponding
to the same area $A$. Hence the calculation of the degeneracy $g(E_n)$ becomes to an easy excercise of
combinatorics: Basically, the calculation of $g(E_n)$ boils down to the question of in how many ways a given positive
integer $n$ may be expressed as a sum of at most $N$ positive integers.

   After finding the allowed values of $E_n$ and the corresponding values of $g(E_n)$ it is easy to write the 
partition function of an acceleration surface. An actual calculation of the partition function, however, is 
rather involved, and the details of that calculation have been expressed in the Appendix A of this paper. 
Nevertheless, the calculation may be carried out explicitly, and the final result turns out to be miraculously
simple. The partition function of an acceleration surface consisting of $N$ Planck size quantum black holes is,
in natural units:
\begin{equation}
Z(\beta) = \frac{1}{2^{\beta T_C}-2}\biggl[1 - \biggl(\frac{1}{2^{\beta T_C} - 1}\biggr)^{N+1}\biggr].
\end{equation}
In this equation
\begin{equation}
T_C := \frac{\alpha\hbar a}{4(\ln 2)\pi k_B c},
\end{equation}
and we call $T_C$ as the {\it characteristic temperature} of an acceleration surface with a proper acceleration $a$.
Eq.(1.8) holds, whenever $\beta T_C \ne 1$. If $\beta T_C = 1$, we have $Z(\beta) = N+1$.

   In Sections 4 and 5 we work out the consequences of Eq.(1.8). In Section 4 we consider the dependence of the
energy $E$ of an acceleration surface on its absolute temperature $T$. It turns out that the characteristic
temperature $T_C$ plays an important role: If $T < T_C$, the energy $E$ is, for large $N$, effectively zero, which 
means that all quantum black holes on an acceleration surface are in vacuum. However, when $T = T_C$, the 
acceleration surface performs a {\it phase transition}, where the energy of the surface increases, although its 
temperature remains the same. During this phase transition the black holes constituting an acceleration surface
jump, in average, from the vacuum to the second excited states. We have investigated this phase transition
both analytically and numerically. When $T > T_C$, the energy of an acceleration surface depends, in effect, 
linearly on its absolute temperature $T$.

  Because the energy of an acceleration surface is effectively zero, when the temperature $T$ measured by an 
observer for the surface is less than its characteristic temperature $T_C$, one may view the temperature $T_C$
as the lowest possible temperature which an accelerated observer may measure; otherwise all black holes on the
surface would be in vacuum. So we find that our model {\it predicts the Unruh effect}. According to the Unruh effect
the Unruh temperature $T_U := \frac{a}{2\pi}$ is the lowest possible temperature which an accelerated observer may 
measure in the sense that $T_U$ is the characteristic temperature of the thermal radiation detected by an accelerated 
observer even when all matter fields are, from the point of view of all inertial observers, in vacuum. 
\cite{yhdeksan} It is a 
common feature of the temperatures $T_C$ and $T_U$ that they are both proportional to the proper acceleration $a$
of the observer. If we identify the temperatures $T_C$ and $T_U$ we may fix the undetermined number $\alpha$ in
Eqs.(1.7) and (1.9). We get:
\begin{equation}
\alpha = 2\ln 2.
\end{equation}
In addition of predicting the Unruh effect, our model also predicts the Hawking effect, which we show explicitly for 
Schwarzschild black holes in Section 4. More precisely, we show that we may take an appropriate acceleration 
surface arbitrarily close to the Schwarzschild horizon of a Schwarzschild black hole, and the temperature measured
by an observer at rest on that surface is exactly the Hawking temperature measured by an observer just outside of
the horizon.

   Section 5 is dedicated to the investigation of the entropic properties of acceleration surfaces. One finds that
for an acceleration surface area $A$ there is a certain critical value $A_{crit}$, which corresponds to the 
situation, where all black holes on the surface are, in average, on the second excited state. If $A < A_{crit}$,
the temperature of the surface is $T_C = T_U$ to a very great precision and the entropy $S$ of the surface is,
for large $N$, directly proportional to $A$. More precisely, the entropy $S$ is, in natural units, {\it 
exactly one-half} of the area, when $N$ goes to infinity. In other words, one gains for the acceleration 
surface entropy 
a value which is exactly {\it twice} the Bekenstein-Hawking entropy of a black hole with the same event horizon 
area. \cite{kymmenen, yksitoista} This result is in harmony with the findings of Refs. 
\cite{kaksitoista, kolmetoista, neljatoista} . When $A > A_{crit}$, however, one finds for 
the entropy $S$ an expression:
\begin{equation}
S(A) = \frac{1}{2\ln 2}\frac{k_B c^3}{\hbar G}A\ln(\frac{2A}{2A - A_{crit}}) 
+ Nk_B\ln(\frac{2A - A_{crit}}{A_{crit}}).
\end{equation}
As one may observe, this expression contains logarithms of the area $A$.

   We close our dicussion in Section 6 with some concluding remarks. Unless otherwise stated, we shall always use 
the natural units, where $\hbar = G = c = k_B = 1$.

 \section{Preliminaries: Acceleration Surfaces and Their Properties}

  \subsection{The Concept of Acceleration Surface}

   During some recent years there has been accumulating evidence that general relativity may be understood in 
terms of the properties of certain spacelike two-surfaces of spacetime. One of the first steps in this direction
was taken by Jacobson already in 1995, when he managed to show that Einstein's field equation may be obtained from
the first law of thermodynamics and an assumption that any finite part of the past Rindler horizon of an 
accelerating observer possesses, in a certain sense, an amount of entropy which, in the natural units, is
exactly one-quarter of the area of that part. \cite{kahdeksan} More precisely, Jacobson 
considered the flow of matter through the
past Rindler horizon of an accelerating observer, and he identified the boost energy flow of the matter through the
horizon as its heat flow. Assuming that the horizon shrinks during the flow of matter through the horizon such that
the amount of entropy carried by the matter through the horizon is, when the temperature of the 
matter equals with the Unruh 
temperature of the observer, exactly one-quarter of the decrease in the horizon area, Jacobson found that Einstein's
field equation is a straightforward consequence of the first law of thermodynamics. Somewhat related investigations
have been made by Padmanabhan and his collaborators. \cite{viisitoista} They have found that Einstein's field equation may be obtained
by varying the boundary term in the Einstein-Hilbert action, when the boundary consists of a horizon of spacetime.

   A horizon of spacetime is a certain null hypersurface, and it is created, when the points of an appropriate
spacelike two-surface move along certain null curves of spacetime. A Rindler horizon in Minkowski spacetime 
equipped with the flat Minkowski metric, for instance, consists of the world lines of the points of the plane
with $x \equiv 0$ for $t = 0$, when the points of that plane move along the null lines, where $x = \pm t$, and the
coordinates $y$ and $z$ of the points are constants. One may therefore wonder, whether Einstein's field
equation could as well be obtained from the properties of a spacelike two-surface whose points move along curves different
from the null curves of spacetime. In other words, is it really necessary to restrict the considerations to
the {\it horizons} of spacetime?

  There really exist spacelike two-surfaces which are not parts of any horizons of spacetime in the sense 
described above, and whose dynamical properties nevertheless imply Einstein's field equation.  A specific 
example of this kind of a surface is the so called {\it acceleration surface}. In broad terms, acceleration
surface may be described as a smooth, orientable, simply connnected, spacelike two-surface of spacetime, whose 
every point is accelerated with a constant proper acceleration $a$ to the direction of a spacelike unit
normal $n^\mu$ of the surface. For instance, a flat plane parallel to the $xy$-plane and accelerating, in the rest
frame of the plane, with a constant proper acceleration to the direction of the $z$-axis in flat Minkowski
spacetime provides a simple example of an acceleration surface. A mathematically precise definition of acceleration
surface is pretty involved, and it deals with the properties of the proper acceleration vector field
\begin{equation}
a^\mu := u^\alpha u^\mu_{;\alpha}
\end{equation}
of a certain smooth congrunce of certain timelike curves. For the sake of brevity and simplicity we shall call the
congruence in question as {\it acceleration congruence}. In Eq.(2.1) the semicolon denotes the covariant derivative,
and $u^\mu$ is the future directed unit tangent vector field of the congruence.

   To be quite precise, acceleration congruence is defined as a smooth congruence of future directed timelike
curves parametrized by the proper time $\tau$ measured along these curves such that:

  (i) All those sets of points, where $\tau = constant$ along the elements of the congruence are smooth, 
orientable, spacelike two-surfaces of spacetime. 

 (ii) The norm, or absolute value
 \begin{equation}
\vert\vert a^\mu\vert\vert := \sqrt{a^\mu a_\mu} := a
\end{equation}
of the proper acceleration vector field $a^\mu$ of the congruence is identically constant.

 (iii) For arbitrary, fixed $\tau$ the proper acceleration vector field $a^\mu$ is parallel to a spacelike
unit normal vector field $n^\mu$ of the spacelike two-surface, where $\tau = constant$.

  (iv) The spacelike two-surface, where $\tau = 0$, intersects orthogonally the elements of the congruence.

  After defining the concept of acceleration congruence we may define {\it acceleration surface}, quite simply,
as an equivalence class of those sets of points, where $\tau = constant$ along the elements of an acceleration
congruence. By definition, the elements of these equivalence classes are smooth, spacelike two-surfaces of 
spacetime. If we pick up any two spacelike two-surfaces of spacetime with these properties, the surfaces are
equivalent, i. e. they belong to the same equivalence class, if they are $\tau = constant$ surfaces of the
{\it same} congruence. In other words, acceleration surfaces are labelled by the corresponding acceleration
congruences. Physically, we may think acceleration surface, as in our heuristic definition, as a certain 
spacelike two-surface propagating in spacetime in a certain way. Viewed in this manner, the acceleration
congruence determining a given acceleration surface constitutes the congruence of the {\it world lines} of the
points of that surface.

   Our definition implies that acceleration surface has a spacelike unit normal vector field $n^\mu$ such that 
\begin{equation}
a^\mu n_\mu \equiv constant := a
\end{equation}
at every point of an acceleration surface propagating in spacetime. So we see that our mathematically precise 
definition reproduces our heuristic definition: All points of an acceleration surface are accelerated with the
same constant proper acceleration $a$ to the direction of a spacelike unit normal vector field $n^\mu$ of the 
surface. It is easy to see that the vector fields $u^\mu$ and $a^\mu$ are orthogonal, i. e. 
\begin{equation}
a^\mu u_\mu \equiv 0,
\end{equation}
and therefore Eq.(2.3) implies:
\begin{equation}
u^\mu n_\mu \equiv 0.
\end{equation}
Eqs.(2.1), (2.3) and (2.5) imply that the vector fields $u^\mu$ and $n^\mu$ will change during the 
propagation of an acceleration surface through space and time such that
\begin{subequations}
\begin{eqnarray}
u^\alpha u^\mu_{;\alpha} &=& an^\mu,\\
u^\alpha n^\mu_{;\alpha} &=& au^\mu.
\end{eqnarray}
\end{subequations}

   An important example of an acceleration surface, in addition to a flat plane accelerating uniformly in flat
Minkowski spacetime, is provided by the equivalence class of the $t = constant$ slices of the timelike hypersurface,
where the Schwarzschild coordinate $r = constant > 2M$ in Schwarzschild spacetime equipped with the Schwarzschild
metric
\begin{equation}
ds^2 = -(1 - \frac{2M}{r})\,dt^2 + \frac{dr^2}{1 - \frac{2M}{r}} + r^2(d\theta^2 + \sin^2\theta\,d\phi^2),
\end{equation}
where $M$ is the Schwarzschild mass. Indeed, the congruence of the timelike, future directed curves, where 
$r$, $\theta$ and $\phi$ are constants constitutes an acceleration congruence: The sets of points, where the 
proper time
\begin{equation}
\tau = (1 - \frac{2M}{r})^{1/2} t
\end{equation}
measured along the elements of the congruence is constant, are spacelike two-spheres with radius $r$, and as such
they are smooth, simply connected, spacelike two-surfaces of spacetime. One also finds that the only non-zero
component of the proper acceleration vector field $a^\mu$ of the congruence in question is
\begin{equation}
a^r = \frac{M}{r^2},
\end{equation}
and therefore its norm
\begin{equation}
\vert\vert a^\mu\vert\vert = \sqrt{a^\mu a_\mu} = (1 - \frac{2M}{r})^{-1/2}\frac{M}{r^2}
\end{equation}
is constant for constant $r$. Since the only non-zero component of the vector field $n^\mu$ is
\begin{equation}
n^r = (1 - \frac{2M}{r})^{1/2},
\end{equation}
we observe that the vectors $a^\mu$ and $n^\mu$ are parallel for every $\tau$ and
\begin{equation}
a^\mu n_\mu = a.
\end{equation}
Hence we have proved that the conditions (i)-(iii) of our definition of acceleration congruence are satisfied.
The condition (iv) holds trivially, because the only non-zero component of the vector field $u^\mu$ is
\begin{equation}
u^t = (1 - \frac{2M}{r})^{-1/2}
\end{equation}
which is orthogonal to the spacelike two-sphere, where $\tau = t = 0$.

     \subsection{Kinematics of Acceleration Surfaces} 

  Acceleration surfaces have many interesting properties, which have been derived in Ref. \cite{kolme}. 
For instance, one may show that for arbitrary $\tau$
an acceleration surface intersects othogonally the world lines of its points. Moreover, the area $A$ of an 
acceleration surface may change when the surface proceeds in space and time. The first proper time derivative
of the area $A$ takes, in general, the form:
\begin{equation}
\frac{dA}{d\tau} = \int_{\mathcal{S}} u^{\mu;\nu}\gamma_{\mu\nu}\,d\mathcal{A},
\end{equation}
where $d\mathcal{A}$ denotes the area element on the acceleration surface $\mathcal{S}$, and the tensor 
$\gamma_{\mu\nu}$ is defined in terms of the fields $u^\mu$ and $n^\mu$ and the spacetime metric $g_{\mu\nu}$
as:
\begin{equation}
\gamma_{\mu\nu} := g_{\mu\nu} + u_\mu u_\nu - n_\mu n_\nu.
\end{equation}
Even more important is the expression to the {\it second} proper time derivative of the area $A$: If the vectors 
$u^\mu$ associated with the points of an acceleration surface are parallel to each other, when $\tau = 0$, i. e. 
\begin{equation}
u^\mu_{;\alpha}E^\alpha_I \equiv 0,
\end{equation}
for arbitrary spacelike, orthonormal tangent vector fields $E^\mu_I$ $(I = 1, 2)$ of the surface, when $\tau = 0$,
the second proper time derivative takes the form:
\begin{equation}
\frac{d^2A}{d\tau^2}\vert_{\tau = 0} = \int_{\mathcal{S}}(ak_n + R_{\mu\nu}u^\mu u^\nu 
- R_{\alpha\mu\nu\beta}n^\alpha n^\beta u^\mu u^\nu)\,d\mathcal{A}.
\end{equation}
$R_{\mu\nu}$ and $R_{\alpha\mu\nu\beta}$, respectively, are the Ricci and the Riemann tensors of spacetime, 
and
\begin{equation}
k_n := n^\mu_{;\nu}E^I_\mu E^\nu_I
\end{equation}
is the trace of the exterior curvature tensor induced on the surface in the direction determined by the vector 
field $n^\mu$. So we see that if the exterior curvature tensor vanishes identically, i. e. 
\begin{equation}
n^\mu_{;\nu}E^\nu_I \equiv 0
\end{equation}
for all $I = 1, 2$ when $\tau = 0$, we have:
\begin{equation}
\frac{d^2A}{d\tau^2}\vert_{\tau = 0} = \int_{\mathcal{S}}(R_{\mu\nu}u^\mu u^\nu 
- R_{\alpha\mu\nu\beta}n^\alpha n^\beta u^\mu u^\nu)\,d\mathcal{A}.
\end{equation}

         \subsection{Thermal Energy of Acceleration Surfaces}

     The primary reason for defining the concept of acceleration surface is that with acceleration surfaces it is
possible to associate a concept somewhat similar to {\it energy}. In general, the concept of energy is very 
problematic in general relativity (See, for instance, the discussion in Ref. \cite{seitseman}.).
 To gain some insight into this problem it is useful to consider the good old 
Newtonian theory of gravitation. A mathematically advanced way of putting Newton's celebrated universal law of
gravitation is to say that the flux of the gravitational field $\vec{g}(\vec{r})$ through a closed, orientable,
non self-intersecting two-surface $\sigma$ of space is proportional to the total mass $M_{tot}$ inside the
surface. More precisely,
\begin{equation}
M_{tot} = -\frac{1}{4\pi G}\oint_\sigma \vec{g}(\vec{r})\bullet\hat{n}(\vec{r})\,dS,
\end{equation} 
where $G$ is Newton's gravitational constant, $\hat{n}(\vec{r})$ is an outward pointing unit normal of the surface, and
$dS$ is its area element. The gravitational field $\vec{g}(\vec{r})$ tells the {\it acceleration} an observer at
the point $\vec{r}$ of space may measure for a test particle in a free fall. Since mass and energy are equivalent,
we may view the right hand side of Eq.(2.21) as the gravitational energy in Newton's theory.

   An interesting aspect of Eq.(2.21) is that the gravitational mass inside a closed two-surface of space may be read
off from the gravitational field $\vec{g}(\vec{r})$ on the surface alone, without any specific knowledge whatsoever
about the gravitational field inside the surface. In other words, if we know the accelerations of test masses in a
free fall at every point of a closed two-surface, we may calculate the mass, and hence the gravitational energy, 
inside the surface. This important observation prompts a natural question: Is it possible to associate, in any 
meaningful way, the concept of energy with accelerating surfaces themselves? After all, for a given total mass 
$M_{tot}$ the right hand side of Eq.(2.21) gives exactly the same result, no matter whether the mass $M_{tot}$
lies in a single point at the centre of the region bounded by the surface, or is uniformly distributed along the
surface itself.

   There are really some indications that it is possible to associate the concept of energy with acceleration 
surfaces. More precisely, it seems that acceleration surface possesses a certain amount of {\it heat}. For 
instance, it follows from general relativistic quantum field theories that an observer at rest with respect to 
an acceleration surface will observe thermal radiation with a characteristic temperature
\begin{equation}
T_U := \frac{a}{2\pi}
\end{equation}
or, in SI units:
\begin{equation}
T_U := \frac{\hbar a}{2\pi k_B c},
\end{equation}
even when all matter fields are, form the point of observers in a free fall, in vacuum. \cite{yhdeksan}
 This effect is known as
the {\it Unruh effect}, and it is one of the most remarkable results of quantum field theory. The temperature 
$T_U$, in turn, is known as the {\it Unruh temperature}, and it may be viewed as the temperature of an acceleration
surface. If acceleration surface possesses a certain temperature, then why should it not possess, at least in some
sense, a certain amount of heat as well?

   It is natural to write the amount of heat possessed by an acceleration surface $\mathcal{S}$ as a 
straightforward relativistic generalization of the right hand side of Eq.(2.21). We just replace the acceleration
$\vec{g}$ of the test particles in a free fall by the proper acceleration $a^\mu$, and the unit normal
$\hat{n}$ by the vector field $n^\mu$. As a result we get a quantity
\begin{equation}
Q_{as} := \frac{1}{4\pi}\int_{\mathcal{S}} a^\mu n_\mu\,d\mathcal{A}.
\end{equation}
Using Eq.(2.3) we find:
\begin{equation}
Q_{as} = \frac{1}{4\pi}aA
\end{equation}
or, in SI units:
\begin{equation}
Q_{as} = \frac{c^2}{4\pi G}aA,
\end{equation}
where $A$ is the area of the acceleration surface. We suggest that $Q_{as}$ gives, at least in certain special 
cases, the heat possessed by an acceleration surface. One should compare Eq.(2.26) with the 
{\it mass formula of black holes}. \cite{kuusitoista} According to that formula the ADM mass of a 
non-rotating black hole in vacuum is
\begin{equation}
M_{ADM} = \frac{1}{4\pi}\kappa A,
\end{equation}
where $A$ is the event horizon area of the hole, and $\kappa$ is the surface gravity at the horizon. $M_{ADM}$ 
gives the maximum  amount of heat, which may be extracted from a black hole when it radiates away. As one may 
observe, the only difference between Eqs.(2.26) and (2.27) is that in Eq.(2.27) we have replaced the proper 
acceleration $a$ of Eq.(2.26) by the surface gravity $\kappa$. The similarity between Eqs.(2.26) and (2.27) 
provides further support for our idea that $Q_{as}$ gives the heat which may be extracted from an acceleration
surface.

  \subsection{Interaction of Acceleration Surfaces with Matter}

   If we accept the view that acceleration surfaces possess a certain amount of heat, we are faced with a 
possibility  that acceleration surface may exchange heat with the matter flowing through the surface. In other
words, the interactions between matter and the geometry of spacetime, which in general relativity explain the
properties of gravity, may actually be some specific, presumably very simple, heat exchange processes
between matter and the acceleration surfaces of spacetime.

   To see how these heat exchange processes may take place, consider a special case, where the acceleration 
surface satisfies the initial conditions (2.16) and (2.19), when $\tau = 0$, and the matter consists of massless,
non-interacting radiation (electromagnetic radiation, for instance) only. In that case one may show that the
boost energy flow
\begin{equation}
\frac{dE_b}{d\tau} = \int_{\mathcal{S}} T_{\mu\nu}n^\mu u^\nu\,d\mathcal{A}
\end{equation}
(boost energy flown during a unit proper time) carried by the radiation through the acceleration surface equals to
its heat flow $\frac{\delta Q_{rad}}{d\tau}$, i. e. 
\begin{equation}
\frac{\delta Q_{rad}}{d\tau} = \frac{dE_b}{d\tau},
\end{equation}
and the first proper time derivative of the area $A$ vanishes, when $\tau = 0$:
\begin{equation}
\frac{dA}{d\tau}\vert_{\tau = 0} = 0.
\end{equation}
In Eq.(2.28) $T_{\mu\nu}$ is the energy momentum stress tensor of the matter fields. Eq.(2.30) implies, through
Eq.(2.26):
\begin{equation}
\frac{\delta Q_{as}}{d\tau}\vert_{\tau = 0} = 0.
\end{equation}
When the radiation flows through the acceleration surface, it interacts with the surface such that both the area
and the heat flow through the surface will change. As a result, the second proper time derivatives of the 
quantities $Q_{as}$ and $Q_{rad}$ may become non-zero, when $\tau = 0$. We postulate for these second proper time
derivatives an equation
\begin{equation}
\frac{\delta^2 Q_{as}}{d\tau^2}\vert_{\tau = 0} = -\frac{\delta^2 Q_{rad}}{d\tau^2}\vert_{\tau = 0}.
\end{equation}
This equation, which describes the heat exchange between radiation and an acceleration surface, will play an
important role in our discussion. Because of that it may be called, in our approach, as the "fundamental equation"
of the thermodynamics of spacetime. Einstein's field equation with a vanishing cosmological constant for general
matter fields (not just radiation) is a straightforward consequence of Eq.(2.32), and it may also be used as a 
derivation of the Unruh and the Hawking effects once after the entropy of an acceleration surface is known. 
\cite{kolme}  

       \subsection{Einstein's Field Equation for Massless, Non-Interacting Radiation}

     When matter consists of massless, non-interacting radiation in thermal equilibrium, it is very easy to obtain
Einstein's field equation from Eq.(2.32). The energy density $\rho$ and the pressure $p$ of such radiation have,
from the point of view of an observer at rest with respect to the acceleration surface, the following properties:
\begin{subequations}
\begin{eqnarray}
\rho &=& T_{\mu\nu}u^\mu u^\nu,\\
   p &=& T_{\mu\nu}n^\mu n^\nu,\\
   p &=& \frac{1}{3}\rho.
\end{eqnarray}
\end{subequations}
It is an important property of the radiation that its energy momentum stress tensor is {\it traceless}, i. e. 
\begin{equation}
T^\alpha_{\,\,\,\alpha} = 0.
\end{equation}
One may show, using Eqs.(2.6) and (2.28), that the rate of change of the boost energy flow through the surface 
is, in general, when $\tau = 0$:
\begin{equation}
\frac{d^2 E_b}{d\tau^2}\vert_{\tau = 0} = a\int_{\mathcal{S}} T_{\mu\nu}(u^\mu u^\nu + n^\mu n^\nu)\,d\mathcal{A}.
\end{equation}
For our radiation field this equals to the rate of change in the heat flow, and using Eq.(2.33) we find:
\begin{equation}
\frac{\delta^2 Q_{rad}}{d\tau^2}\vert_{\tau =0} = \frac{4}{3}a\int_{\mathcal{S}}T_{\mu\nu}u^\mu u^\nu\,d\mathcal{A}.
\end{equation}
In thermal equilibrium the radiation field is homogenous and isotropic. As a consequence, spacetime
expands and contracts in exactly the same ways in all spatial directions, and we have:
\begin{equation}
R_{\alpha\mu\nu\beta}E^\alpha_{(1)}E^\beta_{(1)} = R_{\alpha\mu\nu\beta}E^\alpha_{(2)}E^\beta_{(2)} 
= R_{\alpha\mu\nu\beta}n^\alpha n^\beta
\end{equation}
everywhere on the acceleration surface. Hence Eq.(2.20) implies:
\begin{equation}
\frac{d^2 A}{d\tau^2}\vert_{\tau=0} = \frac{2}{3}\int_{\mathcal{S}}R_{\mu\nu}u^\mu u^\nu\,d\mathcal{A},
\end{equation}
provided that the initial conditions (2.16) and (2.19) are satisfied. Using Eqs.(2.26) and (2.36) we find that
the fundamental equation (2.32) takes the form:
\begin{equation}
\frac{a}{6\pi}\int_{\mathcal{S}}R_{\mu\nu}u^\mu u^\nu\,d\mathcal{A} 
= -\frac{4}{3}a\int_{\mathcal{S}}T_{\mu\nu}u^\mu u^\nu\,d\mathcal{A}.
\end{equation}
Since the acceleration surface $\mathcal{S}$, as well as the vector field $u^\mu$ are arbitrary, we get:
\begin{equation}
R_{\mu\nu} = -8\pi T_{\mu\nu}, 
\end{equation}
which is exactly Einstein's field equation
\begin{equation}
R_{\mu\nu} = -8\pi(T_{\mu\nu} - \frac{1}{2}g_{\mu\nu}T^\alpha_{\,\,\,\alpha}),
\end{equation}
or
\begin{equation}
R_{\mu\nu} - \frac{1}{2}Rg_{\mu\nu} = -8\pi T_{\mu\nu}
\end{equation}
in the special case, where the tensor $T_{\mu\nu}$ is traceless, i. e. Eq.(2.34) holds.

\subsection{Einstein's Field Equation for General Matter Fields}

     As we saw, the fundamental equation (2.32) indeed implies Einstein's field equation for massless, 
non-interacting radiation in thermal equilibrium. For general matter fields the key idea in the derivation of
Einstein's field equation from the fundamental equation is an observation that when an acceleration surface moves
with respect to the matter fields with a velocity very close to that of light, all matter behaves, in the rest
frame of the acceleration surface, in the same way as does massless, non-interacting radiation. 
\cite{nelja} More precisely, 
in the rest frame of an acceleration surface moving with an enormous speed with respect to the matter fields
the components of the tensor $T_{\mu\nu}$ are, in effect, related to each other in the same way as they are for
massless non-interacting radiation fields. Moreover, in the high speed limit the rate of change in the boost
energy flow through an acceleration surface is exactly the rate of change in the heat flow, no matter what
kind of matter we happen to have.

    To consider Eq.(2.32) in the high speed limit we Lorentz boost the vector fields $u^\mu$
and $n^\mu$ at every point of our acceleration surface to the direction of the vector $-n^\mu$. More precisely,
we define the new vector fields $u'^\mu$ and $n'^\mu$ in terms of the vector fields $u^\mu$ and $n^\mu$ such that
\begin{subequations}
\begin{eqnarray}
u'^\mu :&=& \frac{1}{2}(\frac{k^\mu}{\sqrt{\epsilon}} + \sqrt{\epsilon}\,l^\mu),\\
n'^\mu :&=& \frac{1}{2}(\frac{k^\mu}{\sqrt{\epsilon}} - \sqrt{\epsilon}\,l^\mu),
\end{eqnarray}
\end{subequations}
and we replace in Eqs.(2.20) and (2.35) the vector fields $u^\mu$ and $n^\mu$ by the vector fields $u'^\mu$
and $n'^\mu$. In Eq.(2.42) the vector fields $k^\mu$ and $l^\mu$ are future directed and null such that
\begin{subequations}
\begin{eqnarray}
k^\mu :&=& u^\mu - n^\mu,\\
l^\mu :&=& u^\mu + n^\mu.
\end{eqnarray}
\end{subequations}
In other words, the vectors $k^\mu$ and $l^\mu$, respectively, span the past and the future Rindler horizons of
our accelerating   surface. The parameter $\epsilon$ has been defined as:
\begin{equation}
\epsilon := \frac{1-v}{1+v},
\end{equation}
where $v$ is the velocity of the boosted frame of reference with  respect to the original frame. As one may see,
$v$ gets close to 1, the speed of light in the natural units, when $\epsilon\rightarrow 0$, and $v$ goes to zero,
when $\epsilon\rightarrow 1$. 

    If we replace the vector fields $u^\mu$ and $n^\mu$ in Eq.(2.20) by the vector fields $u'^\mu$ and $n'^\mu$
defined in Eq.(2.42) we find, using Eq.(2.26) and the symmetry properties of the Riemann and the Ricci tensors: 
\cite{kolme}
\begin{equation}
\frac{\delta^2 Q_{as}}{d\tau^2}\vert_{\tau = 0} 
= \frac{a}{16\pi\epsilon}\int_{\mathcal{S}}R_{\mu\nu}k^\mu k^\nu\,d\mathcal{A} + \mathcal{O}(1),
\end{equation}
where $\mathcal{O}(1)$ denotes the terms, which are of the order $\epsilon^0$, or higher. Eq.(2.35), in turn, 
implies:
\begin{equation}
\frac{\delta^2 Q_{rad}}{d\tau^2}\vert_{\tau = 0} 
= \frac{a}{2\epsilon}\int_{\mathcal{S}}T_{\mu\nu}k^\mu k^\nu\,d\mathcal{A} + \mathcal{O}(\epsilon),
\end{equation}
where $\mathcal{O}(\epsilon)$ denotes the terms, which are of the order $\epsilon^1$, or higher. So we find that
our fundamental equation implies, in the high speed limit, where $\epsilon\rightarrow 0$:
\begin{equation}
\int_{\mathcal{S}} R_{\mu\nu}k^\mu k^\nu\,d\mathcal{A} 
= -8\pi\int_{\mathcal{S}}T_{\mu\nu}k^\mu k^\nu\,d\mathcal{A}.
\end{equation}
Since the acceleration surface $\mathcal{S}$, as well as the null vector $k^\mu$ are arbitrary, we must have
\begin{equation}
R_{\mu\nu} + fg_{\mu\nu} = -8\pi T_{\mu\nu}
\end{equation}
for some function $f$ of the spacetime coordinates. Using the Bianchi identity
\begin{equation}
(R^\mu_{\,\,\,\nu} - \frac{1}{2}\delta^\mu_\nu R)_{;\mu} = 0,
\end{equation}
we observe that
\begin{equation}
f = -\frac{1}{2}R + \Lambda
\end{equation}
for some constant $\Lambda$, and therefore we arrive at an equation
\begin{equation}
R_{\mu\nu} - \frac{1}{2}Rg_{\mu\nu} + \Lambda g_{\mu\nu} = -8\pi T_{\mu\nu},
\end{equation}
which is Einstein's field equation with the cosmological constant $\Lambda$. So we see that Einstein's field equation
indeed follows, not only for radiation, but for general matter fields, from our fundamental equation (2.32).

   There is an interesting difference between Eq.(2.42), which was obtained from our fundamental equation in the
special case, where matter consists of massless, non-intercating radiation in thermal equilibium only, and 
Eq.(2.52), which was obtained for general matter fields: Eq.(2.52) involves an arbitrary cosmological constant
$\Lambda$, whereas Eq.(2.42) does not. Since Eq.(2.42) is a special case of Eq.(2.52), Eq.(2.52) should reduce
to Eq.(2.42), when matter consists of massless, non-interacting radiation only. Obviously, this is not possible,
unless the cosmological constant will vanish. In other words, our fundamental equation implies that  
we must have:
\begin{equation}
\Lambda = 0.
\end{equation}
So we see that our fundamental equation makes a precise prediction, which is consistent with the current 
observations, which imply that the cosmological constant, although not necessarily exactly zero, must nevertheless
be very small. \cite{seitsemantoista} 

\section{Partition Function of Spacetime}

\subsection{Systems in Gravitational Physics}

     As we saw in the previous Section, Einstein's field equation with a vanishing cosmological constant may be
obtained by means of very simple considerations concerning the thermodynamical properties of spacetime and matter 
fields. We introduced the concept of acceleration surface, associated acceleration surfaces with the concept of
heat, and postulated an equation which tells in which way an acceleration surface and the radiation flowing through
the surface exchange heat. Einstein's field equation was a simple and straightforward consequence of that equation.

    Ever since the works of Boltzmann and, in fact, those of Daniel Bernoulli, who was the first to be able to show
that Boyle's law may be obtained by assuming that gases consist of tiny particles, 
\cite{kahdeksantoista} we have learned that the 
thermodynamical properties of any system follow from the physical properties of its constituents. For instance, we
may calculate the specific heat of a piece of a metal, if we know the characteristic frequencies of the 
oscillations
performed by its atoms in the metallic lattice. The fundamental object in the derivation of the thermodynamical
properties of any system from its microphysics is the {\it partition function}
\begin{equation}
Z(\beta) := \sum_n g(E_n)e^{-\beta E_n}
\end{equation}
of the system. If we know the partition function of a system, we may calculate all of its thermodynamical 
properties. In Eq.(3.1) $\beta$ is, in natural units, the inverse of the absolute temperature of the system,
$n$ is an index which labels the possible energy eigenvalues $E_n$ of the system, and $g(E_n)$ is the degeneracy of
a state with energy $E_n$. In other words, $g(E_n)$ tells the number of the microscopic states of the system 
corresponding to the same total energy $E_n$.

    When attempting to write the partition function of spacetime, one is faced with several questions of a 
fundamental nature: What actually is the system we should investigate? What are the microscopic states of the 
system? What are its microscopic constituents? What is the number of microscopic states corresponding to the same 
total energy of the system?

     We begin with with an investigation of the concept of "system" in gravitational physics. Even if we 
restricted our attention to classical general relativity, there are several possible choices for systems in gravitational
physics. A field theorist, for instance, might say that the system one should investigate in gravitational 
physics is the gravitational field $h_{\mu\nu}$, which may be understood as a small perturbation in the flat
Minkowski metric $\eta_{\mu\nu}$, when we write the spacetime metric $g_{\mu\nu}$ as 
$g_{\mu\nu} = \eta_{\mu\nu} + h_{\mu\nu}$. In contrast, an enthusiastic of canonical gravity would maintain that
it is not the gravitational field but those spacelike hypersurfaces of spacetime, where the time parameter 
$t = constant$, which play the role of systems in gravitational physics. The concept of system becomes 
even more diversed and
more complicated, if one attempts to quantize gravity. For instance, in loop quantum gravity the notion of system 
is totally different from that in string theory.

    One of the defining ideas of this paper is to take, at least in the macroscopic level, {\it acceleration
surfaces} as the systems under consideration. This kind of a choice is quite natural, because we saw in the
previous Section that Einstein's field equation, and thereby the whole classical general relativity with all
of its consequences, may be reduced to the properties of acceleration surfaces in a very simple and straightforward
manner. One of the advantages of taking acceleration surfaces as the systems in gravitational physics is that it
allows one to associate the concept of energy with the physical systems uner study: In this paper we identify the
energy $E$ of an acceleration surface simply as the heat, or thermal energy $Q_{as}$ of Eq.(2.26).

\subsection{Microscopic Properties of Acceleration Surfaces}

    A much more difficult problem is posed by the microscopic structure of our systems. Following the ideas
presented in Refs. \cite{kaksi} and \cite{kolme} we model spacetime by a specific {\it graph}, 
where {\it Planck size quantum black
holes} lie on the vertices. At this point it is not necessary to go into the details of this model. An interested
reader may consult the Refs. \cite{kaksi} and \cite{kolme}. It is sufficient to say that the only physical degree of freedom
associated with a Planck size quantum black hole acting as a fundamental building block of spacetime is
its horizon area, and acceleration surfaces, as well as spacetime as a whole is made, in our model, of these
black holes. The key idea is to reduce all geometrical properties of spacetime to the horizon area eigenvalues
of the black holes, which are assumed to be of the form:
\begin{equation}
A_n = (n + \frac{1}{2})32\pi\ell_{Pl}^2,
\end{equation}
where the possible values of $n$ are 0, 1, 2, 3,..., and $\ell_{Pl}$ is the Planck length. In other words, we 
assume that the quantum black holes have an equal spacing in their horizon area spectrum. An equally spaced horizon
area spectrum for quantum black holes was originally proposed by Bekenstein already in 1974. \cite{viisi} Since then 
Bekenstein's proposal has been recovered by several authors on various grounds. \cite{kuusi}

   The area of an acceleration surface depends on the horizon areas of the Planck size holes constituting that
surface. As in Ref. \cite{kaksi} we pose the following {\it independence postulates} for these holes:

   (IP1) The quantum states of the microscopic quantum black holes constituting an acceleration surface 
$\mathcal{S}$ are independent of each other.

   (IP2) The vacuum, or ground states, where $n = 0$, do not contribute to the area of $\mathcal{S}$. 

   (IP3) When a hole is in the $n$'th excited state, it contributes to $\mathcal{S}$ an area, which is directly
proprotional to $n$.

   (IP4) The total area $A$ of $\mathcal{S}$ is the sum of the areas contributed by the black holes on 
$\mathcal{S}$ to the total area of $\mathcal{S}$.

   The physical meaning of these postulates has been considered in details in Ref. \cite{kaksi}. When put in together, the 
postulates (IP1)-(IP4) imply that the possible eigenvalues of the total area $A$ of an acceleration surface
$\mathcal{S}$ are, in natural units, of the form:
\begin{equation}
A = \alpha(n_1 + n_2 +...+n_N),
\end{equation}
where the non-negative integers $n_1, n_2,..., n_N$ are the quantum numbers labelling the horizon area eigenvalues
of the $N$ microscopic black holes lying on $\mathcal{S}$, and $\alpha$ is a pure number to be determined later.
Since we have decided to identify the total energy $E$ of an acceleration surface, from the point of view of
of an observer at rest with respect to that surface, as the thermal energy $Q_{as}$ of Eq.(2.26), we find that
the eigenvalues of $E$ are, in natural units, of the form:
\begin{equation}
E_n = n\frac{\alpha a}{4\pi}
\end{equation}
or, in SI units:
\begin{equation}
E_n = n\frac{\alpha\hbar a}{4\pi c},
\end{equation}
where $a$ is the proper acceleration of the surface, and $n$ is a non-negative integer such that:
\begin{equation}
n := n_1 + n_2 +...+n_N.
\end{equation}

   In addition to the independence postulates (IP1)-(IP4) the holes constituting $\mathcal{S}$ are assumed to
obey the following {\it statistical postulates}, which specify the micro- snd the macrostates of an acceleration 
surface:

   (SP1) The microstates of an acceleration surface $\mathcal{S}$ are uniquely determined by the combinations of 
the non-vacuum horizon area eigenstates of the quantum black holes on $\mathcal{S}$.

   (SP2) Each  microstate on $\mathcal{S}$ yielding the same energy eigenvalue of $\mathcal{S}$ corresponds to the
same macrostate of $\mathcal{S}$.

     \subsection{Degeneracy of the Energy Eigenvalues}

   Our statistical postulates (SP1) and (SP2), together with the four indpendence postulates (IP1)-(IP4), enable
us to calculate the degeneracy $g(E_n)$ of a given energy eigenvalue $E_n$ of an acceleration surface, when 
$n=1,2,...$. $g(E_n)$ is simply the number of ways to express the positive integer $n$ as a sum of at most $N$ 
positive integers. More precisely, $g(E_n)$ is the number of the ordered strings $(n_1,n_2,...,n_m)$, where 
$n_1, n_2,..., n_m \in \lbrace 1, 2, 3,...,\rbrace$ and $1 \le m\le N$ such that
\begin{equation}
n_1 + n_2 +... + n_m = n.
\end{equation}

   It is pretty easy to find an explicit expression for $g(E_n)$. To begin with, we observe that the number of 
ways of writing a given positive integer $n$ as a sum of exactly $m$ positive integers is, when $m \le n$, given
by the binomial coefficient
\begin{equation}
\Omega_m(n) = \left(\begin{array}{ccc}
              n-1\\
              m-1
                     \end{array}\right).
\end{equation}
For instance, there are 
$\left(\begin{array}{ccc}5-1\\3-1\end{array}\right) = \left(\begin{array}{ccc}4\\2\end{array}\right) = 6$ ways to express a 
number 5 as a sum of exactly 3 positive integers. Indeed, we have:
\begin{equation}
5 = 3 + 1 + 1 = 1 + 3 + 1 = 1 + 1 + 3 = 1 + 2 + 2 = 2 + 1 + 2 = 2 + 2 + 1.
\end{equation}
To see how Eq.(3.8) comes out, consider $n$ identical balls in a row. It is easy to see that $\Omega_m(n)$ is the
number of ways of arranging the $n$ balls in $m$ groups by putting $m-1$ divisions in the $n-1$ available empty
spaces between the balls. There are $\left(\begin{array}{ccc}n-1\\m-1\end{array}\right)$ ways of picking up $m-1$
places for the divisions, and so Eq.(3.8) follows.

    The calculation of $g(E_n)$ is based on Eq.(3.8). Let us first assume that $N$, the number of microscopic black
holes on the acceleration surface $\mathcal{S}$, is smaller than $n$. In that case
\begin{equation}
g(E_n) = \sum_{m=1}^N \Omega_m(n) = \sum_{k=0}^{N-1}\left(\begin{array}{ccc}n-1\\k\end{array}\right).
\end{equation}
In the special case, where $N = n$, we have:
\begin{equation}
g(E_n) = \sum_{k=0}^{n-1}\left(\begin{array}{ccc}n-1\\k\end{array}\right) = 2^{n-1}.
\end{equation}
If $N > n$, $g(E_n)$ is simply the number of ways of expressing $n$ as a sum of positive integers, no matter how 
many. Since the maximum number of those positive integers is $n$, we find that $g(E_n)$ is given by Eq.(3.11),
whenever $N \ge n$.

      \subsection{The Partition Function}

    Our considerations allow us to write the partition function $Z(\beta)$ of Eq.(3.1) for an acceleration surface 
with a proper acceleration $a$. Using Eq.(3.5) for $E_n$ and Eqs.(3.10) and (3.11) for $g(E_n)$ we find:
\begin{equation}
Z(\beta) = Z_1(\beta) + Z_2(\beta),
\end{equation}
where
\begin{subequations}
\begin{eqnarray}
Z_1(\beta) :&=& \sum_{n=1}^N 2^{n-1}e^{-n\beta E_1},\\
Z_2(\beta) :&=& \sum_{N+1}^\infty\biggl[\sum_{k=0}^N\left(\begin{array}{ccc}n-1\\k\end{array}\right)
e^{-n\beta E_1}\biggr].
\end{eqnarray}
\end{subequations}
It turns out useful to define the temperature
\begin{equation}
T_C := \frac{\alpha\hbar a}{4(\ln2)\pi k_B c}.
\end{equation}
When written in terms of $T_C$, $Z_1(\beta)$ and $Z_2(\beta)$ take, in the natural units, the forms:
\begin{subequations}
\begin{eqnarray}
Z_1(\beta) &=& \frac{1}{2}\sum_{n=1}^N 2^{(1-\beta T_C)n},\\
Z_2(\beta) &=& \sum_{n=N+1}^\infty\biggl[\sum_{k=0}^N 
\left(\begin{array}{ccc}n-1\\k\end{array}\right)2^{-n\beta T_C}\biggr].
\end{eqnarray}
\end{subequations}
We shall see later that the temperature $T_C$ will play an important role in the satistical and the thermodynamical 
considerations of our model. We shall call $T_C$ as the {\it characteristic temperature} of an acceleration 
surface.

     The calculation of the partition function has been performed in details in the Appendix A. It is most 
gratifying that the calculations may be performed analytically from the beginning to the end. The final result
turns out to be surprisingly simple. We find:
\begin{equation}
Z(\beta) = \frac{1}{2^{\beta T_C} - 2}\biggl[1 - \biggl(\frac{1}{2^{\beta T_C} - 1}\biggr)^{N+1}\biggr],
\end{equation}
if $\beta T_C \ne 1$, and
\begin{equation}
Z(\beta) = N+1,
\end{equation}
if $\beta T_C = 1$.

\section{Energy of an Acceleration Surface}

    After finding, in Eq.(3.16), an explicit and miraculously simple expression for the partition function of an
arbitrary acceleration surface, we are now able to work out its physical consequences.

    First, let us consider the dependence of the energy $E$ of an acceleration surface on the absolute temperature
$T$ of the surface, when the surface is in thermal equilibrium with its surroundings. If we know the partition
function $Z(\beta)$ of any system, the total energy of the system corresponding to its inverse temperature $\beta$
is, in general,
\begin{equation}
E(\beta) = -\frac{\partial}{\partial\beta}\ln Z(\beta).
\end{equation}
Using Eq.(3.16) we get for the energy $E(\beta)$ of an acceleration surface, in natural units:
\begin{equation}
E(\beta) = \biggl[\frac{2^{\beta T_C}}{2^{\beta T_C} - 2} - \frac{(N+1)2^{\beta T_C}}
{(2^{\beta T_C} - 1)^{N + 2} - 2^{\beta T_C} + 1}\biggr](\ln 2)T_C.
\end{equation}
The number $N$, which tells the number of the Planck size black holes on the acceleration surface under 
consideration, is assumed to be very large. For instance, if the area of the acceleration surface is, 
say, $1m^2$, then $N$ is of the order $10^{70}$. It is therefore useful to divide the quantity $E(\beta)$
by $N$, and to consider the quantity
\begin{equation}
{\bar E}(\beta) := \frac{E(\beta)}{N},
\end{equation}
which tells the energy of an acceleration surface per a hole. Although it makes no physical sense to associate 
the concept of energy with an individual Planck size black hole lying on an acceleration surface, Eq.(4.3) 
nevertheless tells how far the black holes are, in average, from the vacuum. More precisely, the average value of 
the quantum number $n$ of Eq.(3.2) associated with an individual black hole is, in SI units:
\begin{equation}
{\bar n}(\beta) = \frac{1}{\alpha}\frac{4\pi c}{a\hbar}{\bar E}(\beta),
\end{equation}
which follows from Eq.(3.5)

\subsection{The Unruh Effect}

   Eq.(4.2) implies, for very large $N$:
\begin{equation}
{\bar E}(\beta) = \frac{1}{N}\frac{2^{\beta T_C}}{2^{\beta T_C} - 2}(\ln 2)T_C - \frac{2^{\beta T_C}}
{(2^{\beta T_C} - 1)^{N+2} - 2^{\beta T_C} + 1}(\ln 2)T_C.
\end{equation}
When obtaining Eq.(4.5) we have approximated $N+1$ by $N$. Since $N$ is assumed to be very large, we find that,
except for the special case, where $T = T_C$, the first term on the right hand side of Eq.(4.5) becomes negligible,
when compared to the second term. Hence we may write, in effect,
\begin{equation}
{\bar E}(\beta) = \frac{2^{\beta T_C}}{2^{\beta T_C} - 1 - (2^{\beta T_C} - 1)^{N+2}}(\ln 2)T_C,
\end{equation}
whenever $T \ne T_C$. As one may observe, the right hand side of Eq.(4.6) is positive, whenever 
$2^{\beta T_C} - 1 < 1$, which means that $T > T_C$. On the other hand, if $T < T_C$, $2^{\beta T_C} - 1 > 1$, 
and $(2^{\beta T_C} - 1)^{N+2}$ becomes huge for large $N$. Hence we get an important result:
\begin{equation}
\lim_{N\rightarrow\infty}{\bar E}(\beta) = 0.
\end{equation}
for all $T < T_C$. In other words, all Planck size black holes constituting an acceleration surface are in vacuum,
when $T < T_C$. This means that the characteristic temperature $T_C$ defined in Eq.(3.13) is the lowest possible
temperature which an acceleration surface may have from the point of view of an observer moving along with the
surface. Putting this in another way, we may say that when matter fields are in thermal equilibrium with spacetime,
an accelerated observer will always measure for the matter fields a temperature, which is at least $T_C$. An 
importance of this
result lies in its close relationship with the {\it Unruh effect}, which was briefly mentioned in Section 2 as a 
motivation for our decision to associate the concept of energy with an acceleration surface. \cite{yhdeksan}
According to the
Unruh effect an accelerated observer will always observe thermal radiation with the Unruh temperature $T_U$
of Eq.(2.23), which is proportional to the observer's proper acceleration $a$, whereas we found that the minimum
temperature an accelerated observer may ever measure for the matter in thermal equilibrium with spacetime is given
by the characteristic temperature $T_C$, which is also proportional to $a$. So it appears that our model predicts
the Unruh effect, and we may identify the characteristic temperature $T_C$ as the Unruh temperature $T_U$ measured
by an accelerated observer. In other words, we have:
\begin{equation}
T_C = T_U.
\end{equation}
Using Eqs.(2.23) and (3.13) we find:
\begin{equation}
\frac{\alpha\hbar a}{4(\ln 2)\pi k_B} = \frac{\hbar a}{2\pi k_B c},
\end{equation}
which will fix the undetermined numerical constant $\alpha$ such that
\begin{equation}
\alpha = 2\ln 2.
\end{equation}
Hence it follows, through Eq.(3.3), that the possible area eigenvalues of an acceleration surface are of the
form:
\begin{equation}
A_n = 2n\ell_{Pl}^2\ln 2
\end{equation}
where $n = 0, 1, 2,...$. The same result was obtained, by different methods, in Ref.\cite{kolme}.

\subsection{The Hawking Effect}

   In addition of predicting the Unruh effect, our model also predicts the Hawking effect as well. To see how
the Hawking effect comes out from our model at least for Schwarzschild black holes, let us recall from Section 2
that the equivalence class of the $t = constant$ slices of the timelike hypersurfaces, where $r = constant > 2M$
in Schwarzschild spacetime is an acceleration surface with a proper acceleration $a$ given by Eq.(2.10). Using
Eqs.(3.13) and (4.10) we find that the minimum temperature measured by an observer at rest with respect to such
a surface is, in natural units,
\begin{equation}
T_C = (1 - \frac{2M}{r})^{-1/2}\frac{M}{2\pi r^2}.
\end{equation}
When $r$ gets close to the Schwarzschild radius $R_S = 2M$, our acceleration surface approaches the Schwarzschild
horizon and we may write, just outside of the horizon,
\begin{equation}
T_C = (1 - \frac{2M}{r})^{-1/2}\frac{1}{8\pi M},
\end{equation}
which is exactly the Hawking temperature \cite{yksitoista}
\begin{equation}
T_H := \frac{1}{8\pi M},
\end{equation}
corrected by the red shift factor $(1 - \frac{2M}{r})^{-1/2}$.  So we may conclude that the event horizon of a 
Schwarzschild black hole has a minimum temperature which, from the point of view of a faraway observer at rest
with respect to the hole, is given by the Hawking temperature $T_H$ of Eq.(4.14). In other words, we have inferred
the Hawking effect for Schwarzschild black holes from the properties of our partition function.

\subsection{The High Temperature Limit}

   Our next task is to consider the case, where $T > T_C$. In that case $\beta T_C < 1$, which implies that
$2^{\beta T_C} - 1 < 1$. Hence one observes that when $T > T_C$, the term $(2^{\beta T_C} - 1)^{N+2}$ becomes
negligible for very large $N$, and we may write, in effect:
\begin{equation}
{\bar E}(\beta) = \frac{2^{\beta T_C}}{2^{\beta T_C} - 1}(\ln 2)T_C.
\end{equation}
For very high temperature $T >> T_C$, $\beta T_C$ is very small and we may write $2^{\beta T_C}$ as a Taylor
expansion:
\begin{equation}
2^{\beta T_C} = 1 + \beta T_C(\ln 2) + \mathcal{O}[(\beta T_C)^2],
\end{equation}
where $\mathcal{O}[(\beta T_C)^2]$ denotes the terms, which are of the order $(\beta T_C)^2$, or higher. So we may write
Eq.(4.15) as:
\begin{equation}
{\bar E}(\beta) = \frac{1}{\beta} + \mathcal{O}(1),
\end{equation}
where $\mathcal{O}(1)$ denotes the terms, which are of the order $(\beta T_C)^0$, or higher.
This result implies, together with Eq.(4.3), that when the absolute temperature $T$ of an acceleration surface 
is very much higher than its characteristic temperature $T_C$, its total energy is, in natural units,
\begin{equation}
E(T) = NT
\end{equation}
or, in SI units,
\begin{equation}
E(T) = Nk_B T.
\end{equation}

   Eq.(4.19) may be used as a consistency check of our model. It is a general property of any system that in very
high temperature $T$ its thermal energy is of the form:
\begin{equation}
E = \gamma Nk_B T,
\end{equation}
where $N$ is a the number of the constituents of the system, and $\gamma$ is a pure number which depends on the
number of the independent degrees of freedom of each constituent. For instance, the thermal energy of a piece of
an arbitrary solid is given, for sufficiently high temperatures, by the so called Dulong-Petit law: 
\cite{yhdeksantoista}
\begin{equation}
E = 3Nk_B T,
\end{equation}
where $N$ i the number of the fundamental constituents (atoms or molecules) of the solid. As one may observe
from Eq.(4.19), the general high  temperature property given by Eq.(4.20) for any system is also possessed by an
acceleration surface. So it seems that at least in the high temperature limit our model should give a correct
description of the properties of spacetime.

\subsection{Phase Transition}

   So far we have not investigated what happens to ${\bar E}(\beta)$, when $T$ is very close to the characteristic
temperature $T_C$. It has been shown in the Appendix B that 
\begin{equation}
{\bar E} = T_C\ln 2,
\end{equation}
when $T = T_C$, and that
\begin{equation}
\frac{d{\bar E}}{dT}\vert_{T=T_C} = \frac{1}{6}(\ln 2)^2N + \mathcal{O}(1),
\end{equation}
where $\mathcal{O}(1)$ denotes the terms, which are of the order $N^0$, or less. As one may observe, 
$\frac{d{\bar E}}{dT}\vert_{T=T_C}$ becomes very large for large $N$. This means that ${\bar E}$ increases very
fast as a function of $T$, when we are close to $T_C$. Putting this in another way, close to $T_C$ the temperature
$T$ of an acceleration surface remains practically constant when we increase ${\bar E}$. The physical meaning of 
this result is that there is a {\it phase transition} in spacetime, when the temperature $T$ of an acceleration
surface equals to the characteristic temperarure $T_C$: When $T < T_C$, the energy of an acceleration surface per
a hole is virtually zero, whereas at the point, where $T = T_C$, the energy suddenly jumps, and it gets a certain 
finite value. The latent heat ${\bar L}$ per a hole corresponding to this phase transition may be estimated by 
susbstituting $1/T_C$ for $\beta$ in Eq.(4.15). One finds for the latent heat per a hole:
\begin{equation}
{\bar L} = 2(\ln 2)T_C
\end{equation}
or, in SI units:
\begin{equation}
{\bar L} = 2(\ln 2)k_BT_C.
\end{equation}
Since we have seen that $T_C$ may be identified as the Unruh temperature $T_U$ of an observer moving along with 
the acceleration surface we find, using Eq.(2.23):
\begin{equation}
{\bar L} = \frac{\hbar a}{\pi c}\ln 2.
\end{equation}

     The conclusions drawn by means of the analytical approximations performed above are confirmed by the numerical 
results. In Fig. 1 we have made a plot of the average energy ${\bar E}$ per a hole as a function of the absolute 
temperature $T$, when $N = 100$. When $T < T_C$, ${\bar E}$ is practically zero. However, when $T = T_C$, the curve
${\bar E} = {\bar E}(T)$ becomes practically vertical. When $T$ is slightly bigger than $T_C$, ${\bar E}(T)$ is 
approximately $1.4T_C$, which is about the same as $2(\ln2)T_C$. Finally, the dependence of ${\bar E}(T)$ on $T$
becomes approximately linear, when $T >> T_C$.
\begin{figure}[htb!]
\begin{center} 
\includegraphics{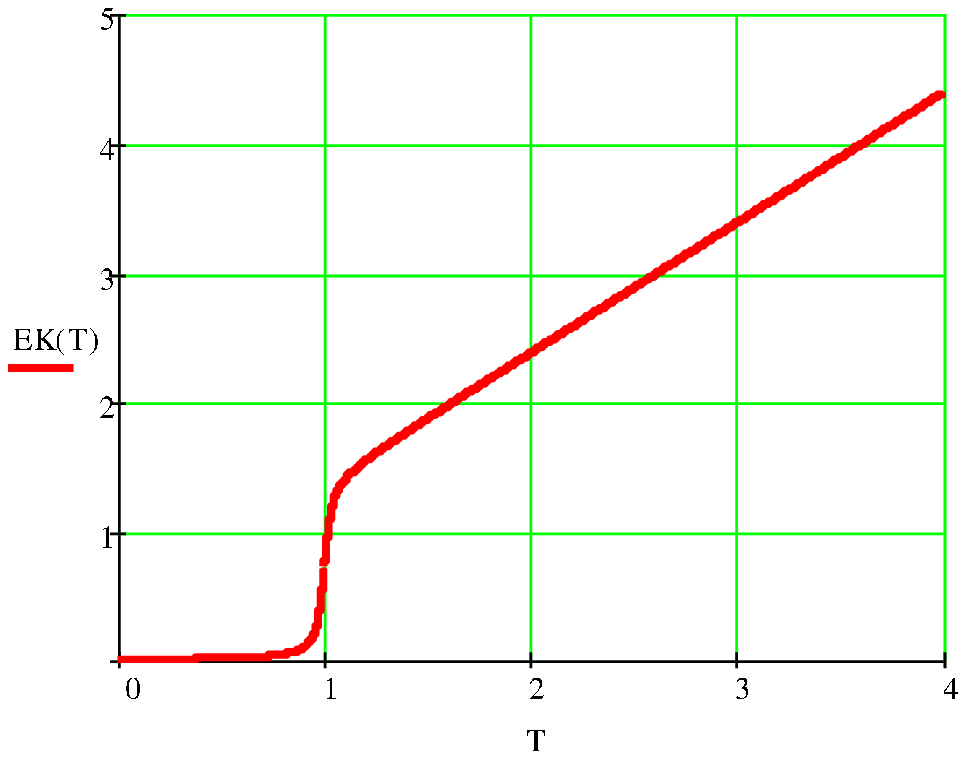}
\caption{The average energy ${\bar E}$ $(= EK(T))$ of an acceleration surface per a hole as a function of the absolute 
temperature $T$, when the number of the holes on the surface is $N = 100$. The absolute temperature $T$ has been 
expressed in the units of $T_C$, and the average energy ${\bar E}$ in the units of $k_BT_C$. If $T < T_C$, 
${\bar E}$ is effectively zero, which means that the black holes on the acceleration surface are in vacuum. When
$T = T_C$, the curve ${\bar E} = {\bar E}(T)$ is practically vertical, which indicates a phase transition at the 
temperature $T = T_C$. During this phase transition the black holes on the acceleration surface are excited from
the vacuum to the second excited states. The latent heat per a hole corresponding to this phase transition is
${\bar L} = 2(\ln 2)k_BT_C \approx 1.4k_BT_C$. When $T > T_C$, the curve ${\bar E} = {\bar E}(T)$ is approximately 
linear.
  \label{fig:Figure1}} 
\end{center}
\end{figure}

     Our analysis shows that the characteristic, or Unruh temperature $T_C$ plays a crucial role in the statistics 
and the thermodynamics of spacetime. The Unruh temperature is the lowest possible temperature an acceleration 
surface may have, and at the Unruh temperature a phase transition occurs with a sudden increase in the energy of 
the acceleration surface.

     It is interesting to consider the microphysical reason for the phase transition observed. What happens to the 
quantum states of the Planck size black holes constituting an acceleration surface during the phase transition? 
Combining Eqs.(4.4) and (4.10) we find that the average value of the quantum number $n$ characterizing the quantum
states of an individual hole on an acceleration surface at the temperature $T = 1/\beta$ is:
\begin{equation}
{\bar n} = \frac{2\pi c}{(\ln 2)a\hbar}{\bar E}(\beta).
\end{equation}
If we substitute for ${\bar E}(\beta)$ the quantity ${\bar L}$ of Eq.(4.26), which gives the average latent heat
per a hole on the surface, we get:
\begin{equation}
{\bar n} = 2.
\end{equation}
The physical meaning of this result is obvious: It means that during the phase transition the Planck size black 
holes on the acceleration surface become excited from the ground state to the second excited state.
In this process the holes absorb quanta of energy from the matter fields until the holes are, in average, in the 
second excited state. The Unruh effect is a process inverse to the excitations of the holes, and it is
caused by the de-excitations of the holes from the second excited state to the ground state.  

\subsection{Zero Point Energy}

     Before closing the discussion about the energy of an acceleration surface it is appropriate to return to 
Eq.(4.2), which gives the precise expression to the total energy $E(\beta)$ of an acceleration surface as a
function of its inverse temperature $\beta$. One finds that although the quantity ${\bar E}(\beta)$ vanishes
for large $N$ and low temperature, the quantity $E(\beta)$ does not: In the low temperature limit $T\rightarrow 0$,
where $\beta T_C\longrightarrow\infty$, the first term inside the brackets on the right hand side of Eq.(4.2)
goes to unity. As a consequence we have:
\begin{equation}
\lim_{T\rightarrow 0}E(\beta) = k_BT_C\ln 2.
\end{equation}
So we see that in our model the total energy of an acceleration surface has a certain non-vanishing zero point
value.

\section{Entropy of an Acceleration Surface}

\subsection{Entropy vs Temperature}

  We shall now turn our attention to the entropy of an acceleration surface. How does the entropy of an acceleration 
surface depend on its temperature? How does it depend on the energy? Finally, we have the most interesting question
of all: How does the entropy of an acceleration surface depend on its area? In particular, what is the relationship
between the acceleration surface entropy and the Bekenstein-Hawking entropy of black holes?

   In general, the entropy $S$ of any system obeys the relationship:
\begin{equation}
F = E - TS,
\end{equation}
where 
\begin{equation}
F := -k_BT\ln Z
\end{equation}
is the Helmholtz free energy of the system. So we find that the entropy of any system may be expressed as a 
function of its inverse temperature $\beta$, in natural units, as:
\begin{equation}
S(\beta) = \beta E(\beta) + \ln Z(\beta).
\end{equation}
Using Eqs.(3.16) and (4.2) we get an expression to the entropy of an acceleration surface:
\begin{equation}
S(\beta) = \biggl[\frac{2^{\beta T_C}}{2^{\beta T_C} - 2} - \frac{(N+1)2^{\beta T_C}}{(2^{\beta T_C} - 1)^(N+2)
- 2^{\beta T_C} + 1}\biggr](\ln 2)\beta T_C + \ln\biggl\lbrace\frac{1}{2^{\beta T_C} - 2}\biggl[1 
- \biggl(\frac{1}{2^{\beta T_C} - 1}\biggr)^{N+1}\biggr]\biggr\rbrace,
\end{equation}
whenever $T \ne T_C$.

  The first observation one may make about the properties of the entropy $S(\beta)$ is that in the limit, where
$T\rightarrow 0$, which means that $\beta \rightarrow\infty$, we have:
\begin{equation}
\lim_{T\rightarrow 0}S(\beta) = 0
\end{equation}
for all $N$. In other words, our system obeys the third law of thermodynamics. Since $N$ is assumed to be of the 
order $10^{70}$, it is useful to consider the rescaled entropy
\begin{equation}
{\bar S}(\beta) := \frac{S(\beta)}{N},
\end{equation}
instead of the entropy $S(\beta)$ itself. One finds that, in general:
\begin{equation}
{\bar S}(\beta) = \frac{1}{N}\frac{2^{\beta T_C}}{2^{\beta T_C} - 2}(\ln 2)\beta T_C - \frac{N+1}{N}
\frac{2^{\beta T_C}}{(2^{\beta T_C} - 1)^{N+2} - 2^{\beta T_C} + 1}(\ln 2)\beta T_C 
+ \frac{1}{N}\ln\biggl\lbrace\frac{1}{2^{\beta T_C} - 2}\biggl[1 
- \biggl(\frac{1}{2^{\beta T_C} - 1}\biggr)^{N+1}\biggr]\biggr\rbrace.
\end{equation}
 
  \subsubsection{The case $T < T_C$}

   Consider first the special case, where $T < T_C$. In that case $2^{\beta T_C} - 1 > 1$, and therefore
$(2^{\beta T_C} - 1)^N$ becomes huge for large $N$. So we observe that
\begin{equation}
\lim_{N\rightarrow\infty}{\bar S}(\beta) = 0,
\end{equation}
whenever $T < T_C$. In other words, the rescaled entropy effectively vanishes for large $N$, when $T < T_C$.
The entropy $S(\beta)$ itself has the property:
\begin{equation}
\lim_{N\rightarrow\infty}S(\beta) = \frac{2^{\beta T_C}}{2^{\beta T_C} - 1}(\ln 2)\beta T_C - \ln(2^{\beta T_C} - 2)
\end{equation}
which, in the natural units, is of the order of unity, when $T < T_C$. We shall see in a moment that 
$S(\beta)$ is of the order of $N$ in the natural units, when $T > T_C$. So we may conclude that not only the does
rescaled entropy ${\bar S}(\beta)$, but also the entropy $S(\beta)$ itself effectively vanishes in the large $N$ 
limit, when $T < T_C$.

  \subsubsection{The case $T > T_C$}

  When $T > T_C$, $0 < 2^{\beta T_C} - 1 < 1$, and $(2^{\beta T_C} - 1)^N$ becomes very small for large $N$.
Hence it follows that when $T > T_C$, we may write for large $N$:
\begin{equation}
{\bar S}(\beta) = \frac{2^{\beta T_C}}{2^{\beta T_C} - 1}(\ln 2)\beta T_C - \ln(2^{\beta T_C} - 1).
\end{equation}
When the temperature is very high, i.e. $T >> T_C$, then $\beta T_C$ is very small, and we find:
\begin{equation}
{\bar S}(\beta) = \ln(\frac{T}{T_C}) - \ln(\ln 2) + \mathcal{O}[(\beta T_C)],
\end{equation}
where $\mathcal{O}[(\beta T_C)]$ denotes the terms, wich are of the order $(\beta T_C)^1$, or higher. Hence
we observe that in very high temperatures the entropy depends, in effect, logarithmically on the temperature $T$.

   \subsubsection{The case $T = T_C$}

   It only remains to consider the case, where $T = T_C$. Using Eqs.(3.17), (4.22) and (5.3) we find:
\begin{equation}
{\bar S}(T_C) = \ln 2 + \frac{1}{N}\ln N.
\end{equation}
Moreover, because 
\begin{equation}
{\bar S}'(T) = \frac{1}{T}{\bar E}'(T),
\end{equation}
which follows from Eqs.(4.1) and (5.3), together with the result:
\begin{equation}
\frac{d\beta}{dT} = -\frac{1}{T^2},
\end{equation}
we find, using Eq.(4.23):
\begin{equation}
{\bar S}'(T_C) = \frac{(\ln 2)^2}{6T_C}N + \mathcal{O}(1),
\end{equation}
where $\mathcal{O}(1)$ denotes the terms, which are of the order $N^0$, or less. So we see that when $T = T_C$,
the first derivative ${\bar S}'(T)$ of ${\bar S}(T)$ is, in effect, proportional to $N$. For large $N$ this 
indicates a rapid jump in the values of ${\bar S}(T)$ at the point, where $T = T_C$. The magnitude of this jump is,
in natural units,
\begin{equation}
\Delta{\bar S} = 2\ln 2,
\end{equation}
which may be obtained from Eq.(5.10), if we substitute $1/T_C$ for $\beta$.

    The conclusions made above are confirmed by numerical investigations. In Fig. 2 we have made a plot of the
graph of the 
function ${\bar S}(T)$ in the special case, where $N = 100$. As one may observe, ${\bar S}(T)$ is virtually zero,
when $T < T_C$. At the point, where $T = T_C$, the curve ${\bar S} = {\bar S}(T)$ is almost vertical, and there 
is a discrete jump $\Delta{\bar S}$, which is about $1.4 \approx 2\ln 2$, in the values of ${\bar S}(T)$. When
$T >> T_C$, ${\bar S}(T)$ depends logarithmically on $T$.
\begin{figure}[htb!]
\begin{center}
\includegraphics{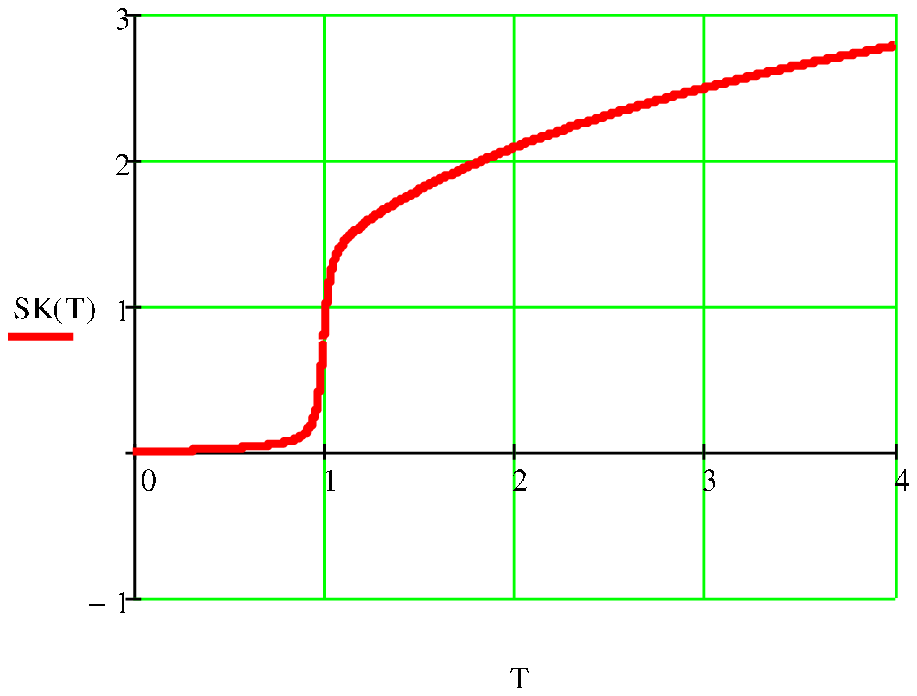}
\caption{The rescaled entropy ${\bar S}$ $(= SK(T))$ (average entropy per a hole) of an acceleration surface as 
a function of the absolute temperature $T$ of spacetime, when the number of black holes on the surface is 
$N = 100$. The absolute temperature
has been expressed in the units of $T_C$, and the rescaled entropy ${\bar S}$ in the units of $k_B$. If $T < T_C$,
${\bar S}$ is effectively zero. At the temperature $T = T_C$ corresponding to the phase transition, the curve
${\bar S} = {\bar S}(T)$ is practically vertical, and there is a discrete jump $\Delta{\bar S} = 2(\ln2)k_B \approx
1.4 k_B$ in the values of ${\bar S}$. Finally, when $T > T_C$, there is an approximately logarithmic dependence 
of ${\bar S}$ on $T$.} 
  \label{fig:Figure2} 
\end{center}
\end{figure}

\subsection{Entropy vs Area}

    We shall now turn our attention to the relationship between the area and the entropy of an acceleration
surface. Our starting point is Eq.(5.13), which implies:
\begin{equation}
\frac{d{\bar S}}{d{\bar E}} = \frac{1}{T}.
\end{equation}

    Consider first the special case, where $T$ is very close to the characteristic temperature $T_C$. In that case
we may write:
\begin{equation}
T = T_C + \Delta T,
\end{equation}
where $\Delta T$, the difference between $T$ and $T_C$, may be approximated by an expression:
\begin{equation}
\Delta T = \frac{dT}{d{\bar E}}\vert_{E=E_C}\,\Delta{\bar E},
\end{equation}
where
\begin{equation}
\Delta{\bar E} := {\bar E} - {\bar E}_C,
\end{equation}
and ${\bar E}_C$ is given by the right hand side of Eq.(4.22), i. e. ${\bar E}_C$ is the average 
energy per a hole 
corresponding to the temperature $T_C$. Because $\frac{dT}{d{\bar E}} = 1/\frac{d{\bar E}}{dT}$ we find, using
Eq.(4.23):
\begin{equation}
\Delta T = \frac{6}{N(\ln 2)^2}\,({\bar E} - {\bar E}_C) + \mathcal{O}(N^{-2}),
\end{equation}
where $\mathcal{O}(N^{-2})$ denotes the terms, which are of the order $N^{-2}$, or less.
Since $N$ is assumed to be very large, we may observe that close to the characteristic temperature $T_C$ the
right hand side of Eq.(5.17) is effectively independent of ${\bar E}$. This is the same conclusion which we may
arrive at, if we look at the curve ${\bar E} = {\bar E}(T)$ of Fig. 1: When $E$ lies within the interval 
$[0, 2T_C\ln 2]$, $T$ is close to $T_C$, and the curve ${\bar E} = {\bar E}(T)$ is practically vertical. This 
means that the temperature $T$ is, in effect, a constant function of ${\bar E}$, when $0 < {\bar E} < 2T_C\ln 2$.
So we find that when $0 < {\bar E} < 2T_C\ln 2$, we may write Eq.(5.17), as an excellent approximation, as:
\begin{equation}
\frac{d{\bar S}}{d{\bar E}} = \frac{1}{T_C}.
\end{equation}
Because the characteristic temperature $T_C$ may identified as the Unruh temperature $T_U$ of Eq.(2.23), and the 
rescaled entropy ${\bar S}$ goes to zero together with the average energy ${\bar E}$, Eq.(5.22) implies, in the
natural units,
\begin{equation}
{\bar S}({\bar E}) = \frac{2\pi}{a}{\bar E}
\end{equation}
or, in the SI units,
\begin{equation}
{\bar S}({\bar E}) = \frac{2\pi k_Bc}{\hbar a}{\bar E},
\end{equation}
whenever $0 < {\bar E} < 2T_C\ln 2$. Multiplying the both sides of Eq.(5.25) by $N$ we get the relationship between
the entropy $S$ and the energy $E$ of an acceleration surface, when $0 < E < 2NT_C\ln 2$:
\begin{equation}
S(E) = \frac{2\pi k_B c}{\hbar a}E.
\end{equation}
If we use Eq.(2.26), which gives the relationship between the area $A$ and the energy $E$ of an acceleration 
surface, we may convert Eq.(5.25) to a relationship between the entropy and the area of an acceleration surface.
In the natural units we get:
\begin{equation}
S(A) = \frac{1}{2}A
\end{equation}
which, in the SI units, takes the form:
\begin{equation}
S(A) = \frac{1}{2}\frac{k_Bc^3}{\hbar G}A.
\end{equation} 

   Eq.(5.27) is most remarkable. It gives an expression for the entropy of an acceleration surface when the surface 
is, from the point of view of an observer at rest with respect to the acceleration surface, in thermal
equilibrium with the matter fields, and the temperature is close
to the Unruh temperature $T_U$ measured by the observer. Eq.(5.27) states that in this case the entropy of an
acceleration surface is, in natural units, exactly {\it one-half} of the area of the surface. Hence we have 
obtained a result which is closely related, although not quite identical, to the Bekenstein-Hawking entropy 
law. According to the Bekenstein-Hawking entropy law the entropy of a black hole event horizon is, in natural
units, one-quarter of its area. \cite{kymmenen, yksitoista} So we have obtained for the acceleration surface entropy an expression, which is
exactly {\it twice} the entropy of a black hole event horizon with the same area.

   The reason for the difference by the factor of two between the right hand side of Eq.(5.27) and that of 
the Bekenstein-Hawking entropy law has been considered in details in Ref. \cite{kolme}. At least for a Schwarzschild black
hole the reason for the difference is easy to understand, if we consider Eq.(2.27), which gives the ADM mass 
$M_{ADM}$ of the hole in terms of the surface gravity $\kappa$ and the area $A$ of its event horizon. 
Differentiating the both sides of Eq.(2.27) we get:
\begin{equation}
dM_{ADM} = \frac{1}{4\pi}A\,d\kappa + \frac{1}{4\pi}\kappa\,dA = \frac{1}{8\pi}\kappa\,dA,
\end{equation}
where we have used the result: \cite{kuusitoista}
\begin{equation}
A\,d\kappa = -\frac{1}{2}\kappa\,dA.
\end{equation}
If we identify $dM_{ADM}$ as the change in the thermal energy of the hole, the fundamental thermodynamical relation
$\delta Q = T\,dS$ implies that $dS = \frac{1}{4}\,dA$, provided that the temperature $T = \frac{\kappa}{2\pi}$. In
contrast, differentiation of the both sides of Eq.(2.25), which gives the thermal energy $Q_{as}$ of an 
acceleration surface, yields the result:
\begin{equation}
\delta Q_{as} = \frac{1}{4\pi}a\,dA,
\end{equation}
which follows from the fact that the proper acceleration $a$ has been kept as a constant during the 
differentiation. Hence the relation $\delta Q = T\,dS$ implies that $dS = \frac{1}{2}\,dA$, when $T = T_U = 
\frac{a}{2\pi}$. So we see that the basic difference between the thermodynamical properties of black hole event
horizons and those of acceleration surfaces is that for black hole event horizons a change in the area implies a 
certain change in its surface gravity, whereas for acceleration surfaces the change in the area preserves the 
proper acceleration $a$, which plays the role of surface gravity for acceleration surfaces, as a constant. It is
this difference, which is the reason for the different entropies of black hole event horizons and acceleration
surfaces.

   One of the crucial points in the derivation of the expression given by Eq.(5.27) for the entropy $S(A)$ of an
acceleration surface, when $T$ is close to $T_C$, was an observation that close to $T_C$ the temperature $T$ is
virtually independent of the energy $E$ of the acceleration surface. However, as we saw in our discussion, an 
independence of the temperature $T$ on the energy $E$ is a valid approximation if and only if ${\bar E}$ is 
smaller than $2T_C\ln 2$, which means that $E$ is smaller than the critical energy
\begin{equation}
E_{crit} := 2NT_C\ln 2 = \frac{Na\ln 2}{\pi}
\end{equation}
or, in SI units:
\begin{equation}
E_{crit} = \frac{N\hbar a\ln 2}{\pi c}.
\end{equation}
If we substitute $E_{crit}$ for $Q_{as}$ in Eq.(2.26), and solve the area $A$ of an acceleration surface, we find
that the critical energy $E_{crit}$ corresponds to the critical area
\begin{equation}
A_{crit} := \frac{4\pi}{a}E_{crit} = 4N\ln 2
\end{equation}
or, in SI units:
\begin{equation}
A_{crit} = 4N\ell_{Pl}^2\ln 2,
\end{equation}
where $\ell_{Pl}$ is the Planck length. The entropy of an acceleration surface is given, as an excellent
approximation, by Eq.(5.27), whenever its area $A$ is less than the critical area $A_{crit}$. When $A = A_{crit}$,
the microscopic black holes are, in average, on the second excited state. 

   What happens, when $A > A_{crit}$? When $A > A_{crit}$, the black holes on the acceleration surface are excited,
in average, above the second excited state, and the temperature exceeds the Unruh temperature $T_U = T_C$. In that
case Eq.(5.27) is no more valid, and we must find another expression for the entropy of an acceleration surface in 
terms of its area $A$.

    Our starting point is Eq.(5.10), which gives the rescaled entropy ${\bar S}(\beta)$ of an acceleration surface
in terms of its inverse temperature $\beta$, when $T > T_C$  and $N$ is very large. As the first step we solve the
quantity $2^{\beta T_C}$ in terms of ${\bar E}$ from Eq.(4.15), which is valid, when $T > T_C$, and 
substitute the resulting expression in Eq.(5.10). We get:
\begin{equation}
{\bar S}({\bar E}) = \frac{{\bar E}}{T_C\ln 2}\ln\biggl(\frac{\bar E}{{\bar E} - T_C\ln 2}\biggr)
+ \ln\biggl(\frac{{\bar E}}{T_C\ln 2} - 1\biggr).
\end{equation}
As one may observe, Eq.(5.35) reduces to Eq.(5.23), when ${\bar E} = 2T_C\ln 2$. So we find that the entropy of 
an acceleration surface may be written in terms of its total energy as:
\begin{equation}
S(E) = \frac{E}{T_C\ln 2}\ln\biggl(\frac{E}{E - NT_C\ln 2}\biggr) 
+ N\ln\biggl(\frac{E - NT_C\ln 2}{NT_C\ln 2}\biggr)
\end{equation}
which, in turn, may be converted to a relationship between entropy and area:
\begin{equation}
S(A) = \frac{1}{2\ln 2}A\ln\biggl(\frac{2A}{2A - A_{crit}}\biggr) 
+ N\ln\biggl(\frac{2A - A_{crit}}{A_{crit}}\bigg)
\end{equation}
or, in SI units:
\begin{equation}
S(A) = \frac{1}{2\ln 2}\frac{k_B c^3}{\hbar G}A\ln\biggl(\frac{2A}{2A - A_{crit}}\biggr)
+ Nk_B\ln\biggl(\frac{2A - A_{crit}}{A_{crit}}\biggr),
\end{equation}
which is valid, whenever $A > A_{crit}$. As one may observe, the entropy is no more a linear function of the
area, but it involves certain logarithmic functions of the area.

   When the area $A$ is close to the critical area $A_{crit}$, the first term on the right hand side of Eq.(5.38)
will dominate. However, when $A >> A_{crit}$, which means that the temperature of the acceleration surface is 
very much higher than its Unruh temperature $T_U$ and the black holes on the surface lie on highly excited states,
the second term will dominate. In this limit we may write, in effect:
\begin{equation}
S(A) = Nk_B\ln(\frac{A}{A_{crit}}),
\end{equation}
where we have neglected the physically irrelevant additive constants from our expression of entropy. In other 
words, in the high temperature limit the entropy of an acceleration surface will no more depend linearly but
logarithmically on its area. This result reproduces the findings of Ref. \cite{kolme}. 

   To conclude, we have found that when $A < A_{crit}$, the entropy of an acceleration surface is, in natural units,
exactly one-half of its area. However, when $A > A_{crit}$, the linear dependence between entropy and area breaks
down, and when $A >> A_{crit}$, the entropy depends logarithmically on the area. These conclusions of ours are
confirmed by numerical investigations. In Fig. 3 we have made a plot of $S$ as a function of $A$, when $N = 100$,
and both $S$ and $A$ have been expressed in the units of $A_{crit}$. One finds that when $A < A_{crit}$, 
$S = \frac{1}{2}A$ as an excellent approximation. However, when $A > A_{crit}$, $S$ will no more depend linearly 
on $A$.
\begin{figure}[htb!]
\begin{center}
\includegraphics{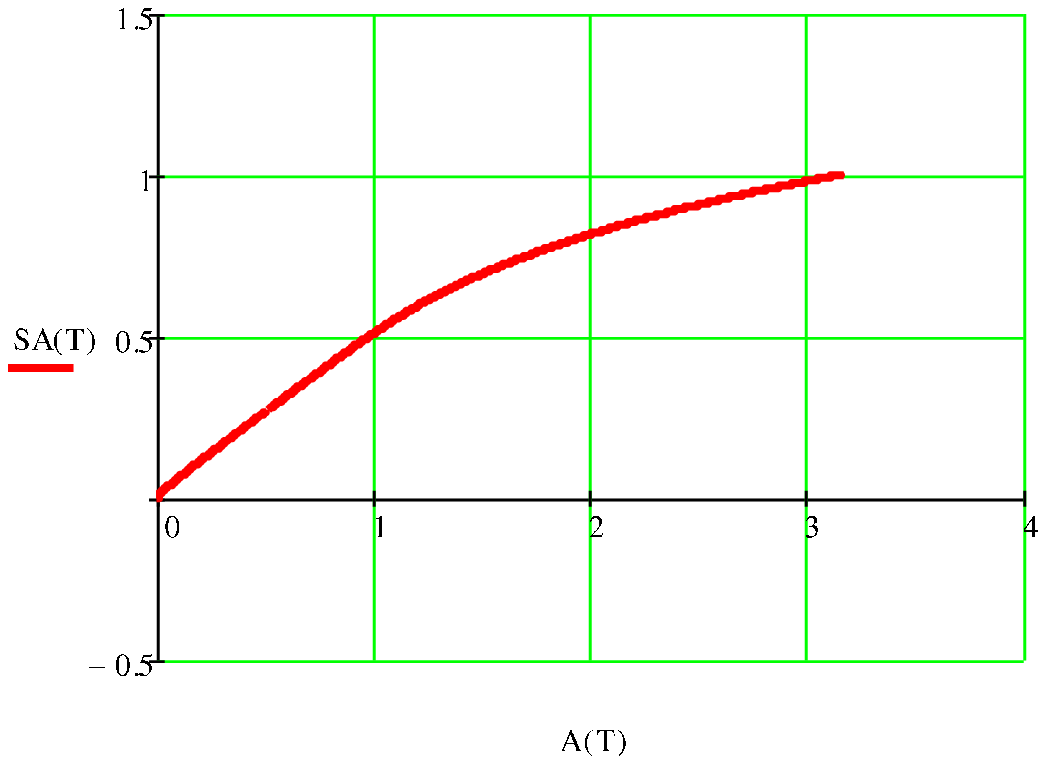}
\caption{The entropy $S$ $(= SA(T))$ of an acceleration surface as a function of the area $A$ of the surface, 
when $N = 100$.
Both $S$ and $A$ have been expressed in the units of the critical area $A_{crit}$. When $A < A_{crit}$, 
$S = \frac{1}{2}A$ as an excellent approximation. However, when $A > A_{crit}$, $S$ will no more depend linearly 
on $A$.} 
  \label{Figure3} 
\end{center}
\end{figure}

\subsection{Equation of State}

  If one knows the entropy $S$ and the temperature $T$ of any system, one may calculate its pressure $p$:
\begin{equation}
p = \biggl(\frac{\partial S}{\partial V}\biggr)_{N, E}T.
\end{equation}
The entropy of a perfect classical gas, for instance, is proportional to the number $N$ of its molecules, and it
depends logarithmically on its volume $V$. Therefore Eq.(5.40) implies for a perfect classical gas the well known 
equation of state:
\begin{equation}
pV = Nk_BT.
\end{equation}

   What is the corresponding equation of state of an acceleration surface? For two-dimensional systems, such as
acceleration surfaces, the pressure $p$ is replaced by the {\it surface tension}
\begin{equation}
\sigma := \biggl(\frac{\partial S}{\partial A}\biggr)_{N,E}T.
\end{equation}
The physical meaning of the surface tension $\sigma$ is that if the area $A$ of the surface is increased by $dA$,
the work done on the surface during the process is
\begin{equation}
dW = \sigma\,dA.
\end{equation}

   Because the work done on an acceleration surface is ultimately converted to heat we find, using Eq.(2.26), that
the surface tension of an acceleration surface with a proper acceleration $a$ is, in general:
\begin{equation}
\sigma_{as} = \frac{c^2 a}{4\pi G}.
\end{equation}
So we see that the surface tension of an acceleration surface depends on its proper acceleration $a$ only, and it 
is  independent of the area $A$ and the temperature $T$ of the surface. The equation of state of an acceleration 
surface gives a relationship between the surface tension $\sigma_{as}$ (and therefore the proper acceleration $a$),
the area and the temperature of the surface. If $A < A_{crit}$, we may use Eq.(5.27), and the resulting equation of 
state  is:
\begin{equation}
\sigma_{as} = \frac{1}{2}T
\end{equation}
or, in SI units:
\begin{equation}
\sigma_{as} = \frac{1}{2}\frac{k_B c^3}{\hbar G}T.
\end{equation}
If $A < A_{crit}$, then $T = T_C = T_U$. If we substitute $T_C$ for $T$ in Eq.(5.47), we recover Eq.(5.45).

    If $A > A_{crit}$, we must use Eq.(5.37). In that case Eq.(5.42) implies the following equation  
of state:
\begin{equation}
\sigma_{as} = \frac{1}{2\ln 2}\frac{k_B c^3}{\hbar G}\biggl[\ln\biggl(\frac{2A}{2A - A_{crit}}\biggr) 
- \frac{A_{crit}}{2A - A_{crit}}\biggr]T + \frac{2Nk_B}{2A - A_{crit}}T.
\end{equation}
In the very high temperatures, where $A >> A_{crit}$, we may write Eq.(5.47), in effect, as:
\begin{equation}
\sigma_{as}A = Nk_BT,
\end{equation}
which is analogous to Eq.(5.41).

\section{Concluding Remarks}

  	In this paper we have considered the partition function of spacetime in gravitational physics. The 
calculation of the partition function was based on a model of spacetime, where spacetime was assumed 
to be a specific graph,
with Planck size quantum black holes on its vertices, and where the macroscopic properties of spacetime were reduced
to the horizon area eigenstates of the holes. In the macroscopic level, the thermodynamical system under 
consideration was taken to be the so called {\it acceleration surface} of spacetime. In broad terms, acceleration
surface may be described as a spacelike two-surface of spacetime, whose every point is accelerated uniformly, with
the same proper acceleration, to the direction of a spacelike unit normal vector field of the surface. For 
acceleration surfaces it is possible to define a quantity which describes the thermal energy of the surface. 
In the classical level, Einstein's field equation with a vanishing cosmological constant
may be shown to be a simple and straightforward consequence of a thermodynamical equation, which describes the 
exchange of energy between an acceleration surface and the matter which flows through the surface, and which 
we called as the "fundamental equation" of the thermodynamics of spacetime. In the quantum level, the Planck size
quantum black holes lying on the acceleration surface were assumed to obey certain independence- and statistical 
postulates. Using these postulates, together with our definition of the concept of thermal energy of an 
acceleration surface, we were able to write the partition function of the surface. We were able to find
an explicit and surprisingly simple expression for the partition function of an acceleration surface in terms
of the inverse temperature $\beta$ of the surface, and to work out the physical consequences of the partition 
function.

    We found that acceleration surface possesses, from the point of view of an observer at rest with respect to
the surface, a certain characteristic temperature $T_C$, which may be identified as the Unruh temperature $T_U$
measured by the observer. When the temperature of an acceleration surface is, from the point of view of our 
observer, less than its Unruh temperature, the energy of the acceleration surface is effectively zero. 
However, when the
temperature $T$ of the surface is the same as its Unruh temperature, the surface performs a {\it phase transition},
where the Planck size quantum black holes on the surface are, in average, excited from the vacuum to the second 
excited  states. During this phase transition the temperature of the surface remains the same, but its energy
jumps from zero to a certain finite, well-defined value. The latent heat corresponding to the 
phase transition equals to the energy which the surface has, from the point of view of our observer, when
all of the Planck size quantum black holes on the surface lie on the second excited state. When the temperature
exceeds the Unruh temperature $T_U$, the energy of the surface depends, in effect, linearly on the temperature 
$T$.

    The same investigations were performed for the entropy of an acceleration surface. As in the case of energy,
it was found that the Unruh temperature $T_U$ of the surface plays an important role. When $T < T_U$, the entropy
of an acceleration surface is effectively zero, but when $T = T_U$, there is a rapid increase in its entropy, 
which corresponds to the phase transition performed by the surface. When $T > T_U$, there is an effective 
logarithmic dependence of the entropy of an acceleration surface on its temperature.

    The entropy $S$ of an acceleration surface may be expressed as a function of the area $A$ of the surface. We 
found that there is a certain critical area $A_{crit}$, which is the area an acceleration surface has, when all
of the Planck size black holes lying on the surface are in the second excited state. When $A < A_{crit}$, the 
entropy of the surface is, in natural units, almost exactly {\it one-half} of the area $A$. However, when 
$A > A_{crit}$, there is no more a linear dependence between the area and the entropy of the surface, but 
certain correction terms involving logarithms of the area will appear. When $A >> A_{crit}$, which means that
the temperature of the acceleration surface vastly exceeds its Unruh temperature and  the black holes are in 
highly excited states, there is, in effect, a logarithmic dependence of the entopy of the surface on its area.

    The most important physical result of this paper is the existence of a phase transition, when the temperature
$T$ of an acceleration surface equals to its characteristic temperature $T_C$. Since the Planck size quantum 
black holes constituting an acceleration surface are in vacuum, when $T < T_C$, and suddenly jump, in average,
to the second excited state, when $T = T_C$, we may view the characteristic temperature $T_C$ as the lowest possible
temperature, which an acceleration surface may have from the point of view of an observer moving along with the 
surface. When acceleration surface is in thermal equilibrium with the radiation fields, it both emits and absorbs
radiation with the temperature $T_C$. As a result, an accelerated observer will detect thermal radiation with a 
characteristic temperature $T_C$, which is proportional to the proper acceleration $a$, even when the radiation
fields are, from the point of view all inertial observers, in vacuum. In other words, our model predicts the 
Unruh effect, and provides that effect with an explanation, which may be traced back to the microscopic
properties of spacetime: In the same way as the thermal radiation of ordinary matter is caused by the 
de-excitations of its atoms and molecules, the thermal radiation observed by an accelerated observer is,
according to our model, caused by the de-excitations of the Planck size quantum black holes constituting
spacetime. The characteristic temperature $T_C$ may be identified as the Unruh temperature $T_U$ measured
by an accelerated observer, and this identification fixed the only undetermined parameter of our model. We 
found in our paper that our model predicts not only the Unruh effect, but also the Hawking effect. That conclusion
was drawn from an observation that every horizon of spacetime, black hole event horizons included, may be 
considered as a limit of an appropriate acceleration surface, when the proper acceleration $a$ of that surface
goes to infinity. 

   To conclude, there were three key elements in our approach to the partition function of spacetime. The first of 
them was our decision to focus our attention to the acceleration surfaces, and to take acceleration surfaces as 
the physical systems under study in gravitational physics. The second element was a definition of the concept of 
energy of an acceleration surface in a certain manner. As we saw in Section 2, that definition implies Einstein's 
field equation  with a vanishing cosmological constant in classical spacetime. Finally, the third element was to 
construct a microscopic model of spacetime out of Planck size quantum black holes, which were assumed to obey 
certain very simple quantum mechanical and statistical postulates. An approach based on these key elements allowed
us to find an explicit and surprisingly simple expression for the partition function of spacetime, and that 
partition function implied, among other things, the Unruh and the Hawking effects. In this sense our model, in
its all simplicity, may hold some promises for the future. It remains to be seen, whether the ideas employed
in this paper in the calculation of the partition function may be utilized in the calculation of the corresponding
quantum mechanical objects, such as the wave function and the propagator, of spacetime.

\appendix

\section{Calculation of the Partition Function}

   In this Appendix derive the expression (3.16) for the partition function $Z(\beta)$ of an acceleration surface
with a proper acceleration $a$. Defining a quantity
\begin{equation}
q := 2^{-\beta T_C}
\end{equation}
one may write the sums $Z_1(\beta)$ and $Z_2(\beta)$ of Eqs.(3.15a) and (3.15b) as:
\begin{subequations}
\begin{eqnarray}
Z_1(\beta) &=& \frac{1}{2}\sum_{n=1}^N (2q)^n,\\
Z_2(\beta) &=& \sum_{n=N+1}^\infty\biggl[\sum_{k=0}^N \left(\begin{array}{ccc}n-1\\k\end{array}\right)q^n\biggr].
\end{eqnarray}
\end{subequations}
Since $\beta$ and $T_C$ are both positive, $q < 1$. $Z_1(\beta)$ is just a geometrical series, and we get:
\begin{equation}
Z_1(\beta) = q\frac{1 - (2q)^N}{1 - 2q},
\end{equation}
provided that $q \ne \frac{1}{2}$. If $q = \frac{1}{2}$, we have 
\begin{equation}
Z_1(\beta) = \frac{1}{2}N.
\end{equation}

     $Z_2(\beta)$ is much more difficult to calculate than $Z_1(\beta)$. When calculating $Z_2(\beta)$, one of the
key ideas is to write the right hand side of Eq.(A2b) by means of the higher order derivatives of an appropriate
function of $q$. Because, in general, an arbitrary binomial coefficient may be written as:
\begin{equation}
\left(\begin{array}{ccc}n\\k\end{array}\right) = \frac{1}{k!}n(n-1)(n-2)...(n-k+1),
\end{equation}
whenever $k > 0$, and $\left(\begin{array}{ccc}n\\k\end{array}\right) = 1$, when $k = 0$, one obtains a general formula
\begin{equation}
\left(\begin{array}{ccc}n\\k\end{array}\right)q^m = \frac{1}{k!}q^{m-n+k}\frac{d^k}{dq^k}q^n,
\end{equation}
which yields:
\begin{equation}
\left(\begin{array}{ccc}n-1\\k\end{array}\right)q^n = \frac{1}{k!}q^{k+1}\frac{d^k}{dq^k}q^{n-1}.
\end{equation}
So we find:
\begin{equation}
Z_2(\beta) = \sum_{n=N+1}^\infty\biggl[\sum_{k=0}^N\frac{1}{k!}q^{k+1}\frac{d^k}{dq^k}q^{n-1}\biggr].
\end{equation}
Since one of the sums has a finite number of terms, we may change the order of the summation, and we get:
\begin{equation}
Z_2(\beta) = \sum_{k=0}^N\biggl[\frac{1}{k!}q^{k+1}\frac{d^k}{dq^k}(q^N\sum_{n=0}^\infty q^n)\biggr].
\end{equation}
Because $\vert q\vert < 1$, the geometric sum on the right hand side of Eq.(A9) will converge, and we have:
\begin{equation}
Z_2(\beta) = \sum_{k=0}^N\biggl[\frac{1}{k!}q^{k+1}\frac{d^k}{dq^k}(\frac{q^N}{1-q})\biggr].
\end{equation}
As one may observe, we have managed to reduce a double sum with an infinite number of terms into a simple sum
with a finite number of terms.

    As the next step we employ the following formula, which is a consequence of the product rule of 
differentiation:
\begin{equation}
\frac{d^k}{dq^k}[f_1(q)f_2(q)] = \sum_{m=0}^k\left(\begin{array}{ccc}k\\m\end{array}\right)f_1^{(k-m)}(q)
f_2^{(m)}(q)
\end{equation}
for arbitrary smooth functions $f_1(q)$ and $f_2(q)$. If we define:
\begin{subequations}
\begin{eqnarray}
f_1(q) &:=& q^N,\\
f_2(q) &:=& \frac{1}{1-q},
\end{eqnarray}
\end{subequations}
we have:
\begin{subequations}
\begin{eqnarray}
f_1^{(k-m)}(q) &=& \frac{N!}{(N-k+m)!}q^{N-k+m},\\
f_2^{(m)}(q) &=& m!(1-q)^{-m-1},
\end{eqnarray}
\end{subequations}
and therefore Eq.(A10) takes the form:
\begin{equation}
Z_2(\beta) = \frac{q^{N+1}}{1-q}\sum_{k=0}^N\biggl[\sum_{m=0}^k\frac{N!}{(k-m)!(N-k+m)!}(\frac{q}{1-q})^m\biggr].
\end{equation}
When obtaining Eq.(A14) we have used the formula:
\begin{equation}
\left(\begin{array}{ccc}k\\m\end{array}\right) = \frac{k!}{m!(k-m)!}.
\end{equation}
Using Eq.(A15) we find:
\begin{equation}
Z_2(\beta) = \frac{x^{N+1}}{(1+x)^N}\sum_{k=0}^N\biggl[\sum_{m=0}^k\left(\begin{array}{ccc}N\\k-m\end{array}\right)
x^m\biggr],
\end{equation}
where we have defined a new variable
\begin{equation}
x := \frac{q}{1-q}.
\end{equation}
Because $0 < q < 1$, $x$ is positive.

   Now, it is possible to rearrange the sums on the right hand side of Eq.(A16). As a result we get:
\begin{equation}
Z_2(\beta) = \frac{x^{N+1}}{(1+x)^N}\sum_{n=0}^N\biggl[\left(\begin{array}{ccc}N\\n\end{array}\right)
\sum_{k=0}^{N-n}x^k\biggr].
\end{equation}
Because 
\begin{equation}
\sum_{k=0}^{N-n}x^k = \frac{1 - x^{N-n+1}}{1-x},
\end{equation}
when $x \ne 1$, and 
\begin{equation}
\sum_{k=0}^{N-n} x^k = N - n + 1,
\end{equation}
when $x = 1$, we have:
\begin{equation}
Z_2(\beta) = \frac{x^{N+1}}{(1+x)^N}\frac{1}{1-x}\biggl[\sum_{n=0}^N\left(\begin{array}{ccc}N\\n\end{array}\right)
- x^{N+1}\sum_{n=0}^N\left(\begin{array}{ccc}N\\n\end{array}\right)x^{-n}\biggr],
\end{equation}
when $x \ne 1$, and
\begin{equation}
Z_2(\beta) = \frac{1}{2^N}\sum_{m=0}^N\biggl[\left(\begin{array}{ccc}N\\n\end{array}\right)(N - n + 1)\biggr],
\end{equation}
when $x = 1$. Using the formulas:
\begin{subequations}
\begin{eqnarray}
\sum_{n=0}^N\left(\begin{array}{ccc}N\\n\end{array}\right) &=& 2^N,\\
\sum_{n=0}^N n\left(\begin{array}{ccc}N\\n\end{array}\right) &=& N2^{N-1},\\
\sum_{n=0}^N\left(\begin{array}{ccc}N\\n\end{array}\right)(\frac{1}{x})^n &=& \biggl(1 + \frac{1}{x}\biggr)^N,
\end{eqnarray}
\end{subequations}
we get:
\begin{equation}
Z_2(\beta) = \frac{x}{1-x}\biggl[(\frac{2x}{1+x})^N - x^{N+1}\biggr],
\end{equation}
when $x \ne 1$, and 
\begin{equation}
Z_2(\beta) = \frac{1}{2}N + 1,
\end{equation}
when $x = 1$. When written in terms of the variable $q$, Eq.(A24) takes the form:
\begin{equation}
Z_2(\beta) = \frac{q^{N+1}}{1 - 2q}\biggl[2^N - \frac{q}{(1-q)^{N+1}}\biggr].
\end{equation}
Combining Eqs.(A2-A4), and (A26) we get, when $\beta \ne \frac{1}{T_C}$:
\begin{equation}
Z(\beta) = \frac{q}{1 - 2q}\biggl[1 - \biggl(\frac{q}{1-q}\biggr)^{N+1}\biggr],
\end{equation}
and
\begin{equation}
Z(\beta) = N + 1,
\end{equation}
when $\beta = \frac{1}{T_C}$. Using Eq.(A1) we find the final expression for the partition function, when 
$\beta \ne  \frac{1}{T_C}$:
\begin{equation}
Z(\beta) = \frac{1}{2^{\beta T_C} - 2}\biggl[1 - \biggl(\frac{1}{2^{\beta T_C} - 1}\biggr)^{N+1}\biggr],
\end{equation}
which is Eq.(3.16).

\section{Properties of the Partition Function Near the Characteristic Temperature}

  In this Appendix we consider the energy of an acceleration surface, when the absolute temperature
$T$ measured by an observer moving along with the surface is very close to the characteristic temperature $T_C$.

   Our starting point is Eq.(3.16), which gives the precise epression for the partition function $Z(\beta)$ of an
acceleration surface. Because $2^{\beta T_C} = 2$, when $T = T_C$, we denote:
\begin{equation}
y := 2^{\beta T_C} - 2,
\end{equation}
and Eq.(3.16) takes the form:
\begin{equation}
Z(y) = \frac{1}{y}[1 - (1+y)^{-N-1}].
\end{equation}
When $T$ is close to $T_C$, $y$ is close to zero. When $y$ is close to zero we may write, using Newton's binomial 
theorem:
\begin{equation}
(1 + y)^{-N - 1} = 1 - (N+1)y + \frac{(N+1)(N+2)}{2!}y^2 - \frac{(N+1)(N+2)(N+3)}{3!}y^3 + ...,
\end{equation}
and we get the Taylor expansion of $Z(y)$ around the point, where $y = 0$:
\begin{equation}
Z(y) = (N+1) - \frac{(N+1)(N+2)}{2!}y + \frac{(N+1)(N+2)(N+3)}{3!}y^2-....
\end{equation}
Applying the chain rule and the result
\begin{equation}
\frac{dy}{d\beta} = T_C(\ln 2)2^{\beta T_C}
\end{equation}
we find:
\begin{equation}
E(\beta) = -\frac{\partial}{\partial\beta}\ln Z(\beta) = -\frac{Z'(y)}{Z(y)}T_C(\ln 2)(y+2),
\end{equation}
where $Z(y)$ is given by Eq.(B4) and
\begin{equation}
Z'(y) = -\frac{(N+1)(N+2)}{2} + \frac{(N+1)(N+2)(N+3)}{3}y - \frac{(N+1)(N+2)(N+3)(N+4)}{8}y^2 +...
\end{equation}
One readily finds that
\begin{subequations}
\begin{eqnarray}
Z(0) &=& N+1,\\
Z'(0) &=& - \frac{(N+1)(N+2)}{2},
\end{eqnarray}
\end{subequations}
which implies:
\begin{equation}
E(\frac{1}{T_C}) = (N+2)T_C\ln 2.
\end{equation}
Therefore, for very large $N$:
\begin{equation}
{\bar E}(\frac{1}{T_C}) = T_C\ln 2,
\end{equation}
which is Eq.(4.22).

   It is interesting to consider the derivative of ${\bar E}$ with respect to $T$, when $T = T_C$. Using Eq.(B6)
we get:
\begin{equation}
\frac{dE}{dy} = \bigg\lbrace\biggl[-\frac{Z''(y)}{Z(y)} + \biggl(\frac{Z'(y)}{Z(y)}\biggr)^2
\biggr](y+2) - \frac{Z'(y)}{Z(y)}\bigg\rbrace T_C\ln 2,
\end{equation}
and Eq.(B4) implies:
\begin{equation}
\frac{dE}{dT}\vert_{T=T_C} = \frac{(\ln 2)^2}{6}(N + 2)(N + 3),
\end{equation}
where we have used the result:
\begin{equation}
\frac{dy}{dT} = -\frac{(\ln 2)T_C}{T^2}2^{\beta T_C}.
\end{equation}
Hence we find that for very large $N$ we may write, in effect:
\begin{equation}
\frac{d{\bar E}}{dT}\vert_{T=T_C} = \frac{(\ln 2)^2}{6}N + \mathcal{O}(1),
\end{equation}
which is Eq.(4.23). $\mathcal{O}(1)$ denotes the terms, which are of the order $N^0$, or less.

\end{document}